\documentclass[pra,
superscriptaddress,
showpacs,preprintnumbers,amsmath,amssymb]{revtex4}

\usepackage{graphicx}
\usepackage[caption=false]{subfig}
\usepackage{booktabs}
\usepackage{diagbox}
\usepackage{amsthm}
\usepackage{tensor}
\usepackage{color}
\usepackage[all]{xy}
\usepackage{tikz}
\usepackage{dsfont}
\usepackage{times,txfonts}
\usetikzlibrary{positioning}
\usepackage{braket}
\usepackage{mathtools}

\newtheorem{Thm}{Theorem}

\newtheorem{Prop}[Thm]{Proposition}

\theoremstyle{definition}
\newtheorem{Def}[Thm]{Definition}

\begin{document}

\title{Improved bounds on quantum uncommon information}

\author{Yonghae Lee} \email{yonghaelee@kangwon.ac.kr}
\affiliation{
Department of Liberal Studies, Kangwon National University, Samcheok 25913, Republic of Korea}

\author{Joonwoo Bae} \email{joonwoo.bae@kaist.ac.kr}
\affiliation{
School of Electrical Engineering, Korea Advanced Institute of Science and Technology (KAIST),
291 Daehak-ro, Yuseong-gu, Daejeon 34141, Republic of Korea}

\author{Hayata Yamasaki} \email{hayata.yamasaki@gmail.com}
\affiliation{
Department of Physics, Graduate School of Science, The University of Tokyo, 7--3--1 Hongo, Bunkyo-ku, Tokyo 113--0033, Japan}

\author{Soojoon Lee}\email{level@khu.ac.kr}
\affiliation{
Department of Mathematics and Research Institute for Basic Sciences, Kyung Hee University, Seoul 02447, Korea}

\pacs{
03.67.Hk, 
89.70.Cf, 
03.67.Mn 
}
\date{\today}

\begin{abstract}
In classical information theory, channel capacity quantifies the maximum number of messages that can be reliably transmitted using shared information.
An equivalent concept, termed {\it uncommon information}, represents the number of messages required to be exchanged to completely share all information in common.
However, this equivalence does not extend to quantum information theory.
Specifically, {\it quantum uncommon information} is operationally defined as the minimal amount of entanglement required for the quantum communication task of {\it quantum state exchange}, where two parties exchange quantum states to share all quantum messages in common.
Currently, an analytical closed-form expression for the quantum uncommon information remains undetermined.
In this work, by investigating underlying characterization of the quantum uncommon information, we derive improved bounds on it.
To obtain these bounds, we develop a {\it subspace exchange strategy} that leverages a common subspace of two parties to identify the unnecessary qubits for exchange.
We also consider a {\it referee-assisted exchange}, wherein a referee aids two parties in efficiently performing the quantum state exchange.
Our bounds provide more precise estimations for the quantum uncommon information.
Furthermore, we demonstrate that the subspace technique is a versatile tool for characterizing uncommon information not only in the bipartite scenario but also in various multi-partite ones.
\end{abstract}

\maketitle

\section{Introduction}
In two-party communication between Alice and Bob, a fundamental problem is to find the maximal rate at which they can reliably transmit information.
In Shannon's information theory, this rate depends on the shared or common information between the parties, denoted by mutual information $I(A;B)$, which defines the channel capacity~\cite{Cover2006}.
Conversely, this also characterizes the {\it uncommon} information, $H(AB) - I(A;B)$, quantifying the amount of classical communication required for them to reliably exchange their messages~\cite{Oppenheim}.
The uncommon information represents the number of bits necessary for the parties to learn the fractions of information that are not in common; technically, Alice and Bob can apply Slepian-Wolf coding~\cite{Slepian1973} to achieve the uncommon information.
The lesson from these operational characterizations is that two-party information can be interpreted as disjoint fractions of common and uncommon information.

In quantum information theory, where messages are in the form of quantum bits or quantum states, the maximal rate at which two parties can reliably transmit quantum states corresponds to quantum common information (QCI), which quantifies the amount of information about a reference system that the parties share in common~\cite{Oppenheim}.
Quantum uncommon information (QUI) is defined as the amount of quantum communication needed to exchange their quantum states, which can be interpreted as the amount of information about a reference system that the parties do not share in common.
Specifically, the quantum state exchange (QSE) task of Alice and Bob defines these quantities~\cite{Oppenheim}.
In this task, Alice and Bob exchange their respective parts, $A$ and $B$, of a quantum state $\psi_{ABR}$ via local operations and classical communication (LOCC) using shared entanglement.
In the asymptotic regime, the QUI of the state $\psi_{ABR}$, denoted by $\Upsilon(A;B)_\psi$, is defined as the minimum rate of pure entanglement required to achieve the QSE task, and the QCI is defined as $S(AB)_\psi - \Upsilon(A;B)_\psi$.

Although the definition of uncommon information is operationally clear in both classical and quantum contexts, characterizing the QUI is not as straightforward as its classical counterpart, and its closed form remains undetermined.
A candidate for this is a seeming quantum analogy, $S(AB)_\psi - I(A;B)_\psi = S(A|B)_\psi + S(B|A)_\psi$.
However, while this analogy can be negative, the QUI must be non-negative~\cite{Oppenheim}.
Therefore, it does not suffice to characterize the QUI.
From an operational standpoint, this indicates that Alice and Bob cannot gain extra entanglement from the QSE task.

Much effort has been devoted to finding a closed form of the QUI.
Ref.~\cite{Lee2019a} allowed Alice and Bob to use quantum side information (QSI), Ref.~\cite{Lee2021} generalized the QSE task to a multi-party communication task, termed quantum state rotation, and Ref.~\cite{Lee2019} investigated the minimal entanglement consumption in an exact single-shot version of the QSE task.
The primary aim of these efforts is to find a closed form of the QUI by considering different variations of the QSE task and determining their minimal entanglement consumption.
Despite these efforts, very little is known about the QUI, and no viable candidates have been proposed.

In this work, we present improved bounds on the QUI by developing brand new techniques called a subspace exchange strategy and a referee-assisted exchange task.
Through the subspace exchange strategy, Alice and Bob can exclude a common state that does not need to be exchanged, focusing solely on the remaining state.
To handle the common state, we formally define a common subspace and the common state within it.
The entanglement rate of this strategy establishes a tight achievable bound on the QUI.
Additionally, with the referee's assistance in our referee-assisted exchange task, the two parties can perform the QSE task more efficiently, thereby providing a tight converse bound on the QUI.

The problem of finding a closed form for non-local resources in uni-directional communication, such as Schumacher compression~\cite{Schumacher1995}, quantum state merging~\cite{Horodecki2005, Horodecki2007}, or quantum state redistribution~\cite{Devetak2008, Yard2009}, has been well studied.
However, despite its theoretical importance, the primitive form of bi-directional communication has seen little progress.
Through our techniques deriving the improved bounds, this study deepens the understanding of the QUI and provides intuition for finding its closed form.
Moreover, these techniques help in reducing the non-local resources required in various quantum communication scenarios, enabling efficient implementations of non-local SWAP and information exchange among three parties~\cite{Lee2021}.

The remainder of this article is organized as follows:
In Sec.~\ref{sec:Defs}, we present the formal descriptions of the QSE task and the QUI, and review the previous bounds on the QUI.
In Sec.~\ref{sec:CommonSubspace}, we define the notion of a common subspace for a tripartite pure state and provide simple examples. 
In Sec.~\ref{sec:SES}, we devise a subspace exchange strategy based on the common subspace, offering an achievable bound on the QUI.
In Sec.~\ref{sec:RAE}, we present a referee-assisted exchange task, providing a converse bound on the QUI.
In Sec.~\ref{sec:Tightness}, we demonstrate the tightness of our bounds with non-trivial examples.
In Sec.~\ref{sec:Applications}, we explain how our techniques can be applied.
Finally, we summarize our results and comment on future works in Sec.~\ref{sec:Conclusion}.

\section{Quantum state exchange and quantum uncommon information} \label{sec:Defs}
Let us begin by reviewing the QSE task and its related information-theoretical quantification, the QUI~\cite{Oppenheim}.
Throughout, we consider finite-dimensional Hilbert spaces. A quantum channel from a Hilbert space $\mathcal{H}_A$ to another Hilbert space $\mathcal{H}_B$ corresponds to a linear, completely positive, and trace-preserving map $\mathcal{N}\colon\mathcal{L}(A)\to\mathcal{L}(B)$~\cite{Wilde2013}, where $\mathcal{L}(X)$ denotes the set of linear operators on the Hilbert space $\mathcal{H}_X$, representing a quantum system $X$.
For clarity, we denote an identity map by $\mathrm{id}_{A}$ and an identity matrix by $\mathds{1}_{A}$ on a quantum system $A$.

\begin{figure}[t]
\includegraphics[clip,width=.75\columnwidth]{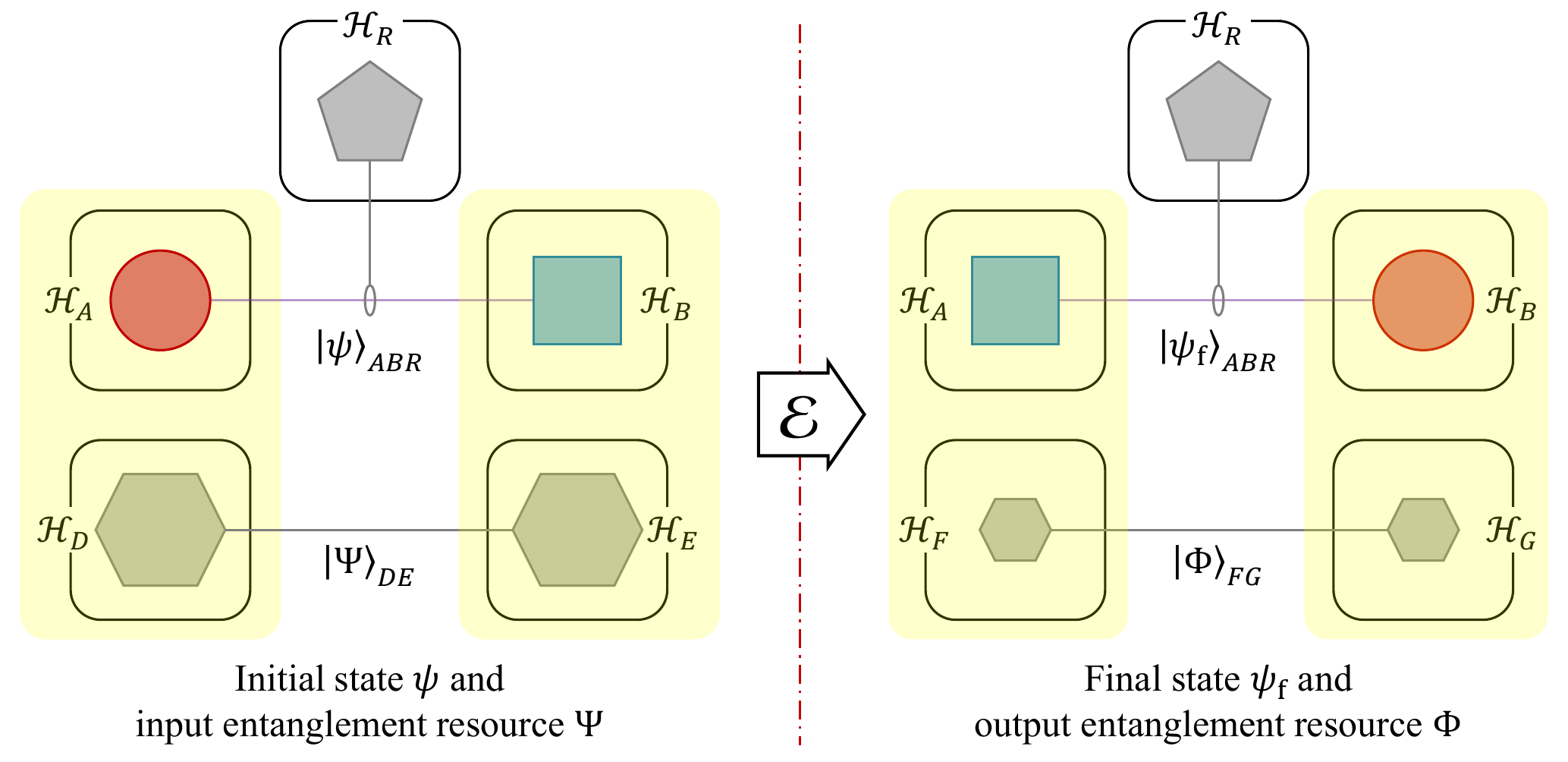}
\caption{
Illustration of QSE protocols:
Basis vectors in Hilbert spaces represent the quantum states of the respective quantum systems.
We visualize them as a circle and polygons, with lines representing correlations among them.
The QSE task involves a two-party scenario where Alice and Bob exchange their information about quantum systems $A$ and $B$ while preserving the correlations between the reference system $R$ and the rest $AB$.
To be specific, a QSE protocol $\mathcal{E}$ in Eq.~(\ref{eq:tr}) transforms an initial state $\psi$ into a final state $\psi_\mathrm{f}$ as shown in Eq.~(\ref{eq:Final}), which is LOCC acting on the shaded area.
The entanglement consumed and gained during the protocol are denoted by $\Psi$ and $\Phi$ in Eq.~(\ref{eq:tr}), respectively.
The amount of entanglement gained from the protocol cannot exceed that of entanglement consumed~\cite{Oppenheim, Lee2019a}.
We express this notion by representing the sizes of the hexagons.
}
\label{fig:QSE}
\end{figure}

To describe a QSE protocol, let us consider a two-party scenario involving Alice and Bob, who may share entanglement and apply LOCC. Let $A$ and $B$ denote quantum systems of Alice and Bob, respectively, whose Hilbert spaces are the same, i.e., $\mathcal{H}_A=\mathcal{H}_B$.
They also have additional systems, $DF$ and $EG$ respectively, to handle entanglement resources.
Then, a quantum channel
\begin{equation}
\mathcal{E} \colon \mathcal{L}\left( AB \otimes DE \right) \longrightarrow \mathcal{L}\left( AB \otimes FG \right)
\end{equation}
is a \emph{QSE protocol} with error $\epsilon$ if it can achieve the following LOCC transformation, as illustrated in Fig.~\ref{fig:QSE},
\begin{equation} \label{eq:tr}
\left\|
\left( \mathcal{E} \otimes \mathrm{id}_{R} \right)
[ \psi \otimes \Psi ] - \psi_\mathrm{f} \otimes \Phi
\right\|_{1}
\le \epsilon,
\end{equation}
where $\|\cdot\|_1$ indicates the trace norm~\cite{Wilde2013}.
The final state $\psi_\mathrm{f}$ of the QSE task is defined from an initial state $\psi$ in such a way that parts $A$ and $B$ are exchanged,
\begin{equation} \label{eq:Final}
\ket{\psi_\mathrm{f}}_{ABR}
\coloneqq \left( \sum_{i,j} \ket{j}\bra{i}_A \otimes \ket{i}\bra{j}_B \right) \otimes \mathds{1}_{R}
\ket{\psi}_{ABR}.
\end{equation}
The symbols $\Psi$ and $\Phi$ denote pure maximally entangled states of the systems $DE$ and $FG$, respectively.

One of the main questions about the QSE task is to find protocols that achieve the task with minimal entanglement in the asymptotic regime.
To be specific, let us consider that two parties have $n$ copies of the initial state $\psi$ and aim to transform them into $n$ copies of the final state $\psi_\mathrm{f}$ using LOCC and shared entanglement.
For each $n$, let $\mathcal{E}_n$ denote a QSE protocol with error $\epsilon_n$,
\begin{equation} \label{eq:EnDomainRange}
\mathcal{E}_n\colon \mathcal{L}\left( A^{\otimes n}B^{\otimes n} \otimes D_nE_n\right) \longrightarrow \mathcal{L}\left( A^{\otimes n}B^{\otimes n} \otimes F_nG_n\right),
\end{equation}
satisfying the following transformation:
\begin{equation} \label{eq:TraceNorm}
\left\|
\left( \mathcal{E}_n \otimes \mathrm{id}_{R^{\otimes n}} \right)
[ \psi^{\otimes n} \otimes \Psi_n ]
- 
\psi_\mathrm{f}^{\otimes n} \otimes \Phi_n
\right\|_{1}
\le \epsilon_n,
\end{equation}
where $\epsilon_n \to 0$ as $n \to \infty$.
In the QSE protocol, $D_nF_n$ and $E_nG_n$ are the quantum systems of Alice and Bob, respectively.
$\Psi_n$ and $\Phi_n$ denote pure maximally entangled states on the quantum systems $D_nE_n$ and $F_nG_n$, respectively.

Through a sequence of such protocols, the amount of entanglement consumed per copy of the initial state is calculated as
\begin{equation}
\frac{1}{n}\left( \log \mathrm{Sr}[\Psi_n] - \log \mathrm{Sr}[\Phi_n] \right),
\end{equation}
where logarithms are taken to base two, and the Schmidt rank of quantum states is used as a measure of entanglement, denoted by $\mathrm{Sr}$.
A real number $r$ is said to be an \emph{achievable} entanglement rate if there exists a sequence of QSE protocols $\mathcal{E}_n$ with errors $\epsilon_n$ such that the following conditions are met:
\begin{eqnarray}
\lim_{n\to\infty} \frac{1}{n}\left( \log \mathrm{Sr}[\Psi_n] - \log \mathrm{Sr}[\Phi_n] \right) &=& r, \\
\lim_{n\to\infty} \epsilon_n &=& 0.
\end{eqnarray}
The \emph{minimal} achievable entanglement rate across all QSE protocols then defines the QUI $\Upsilon$ of a tripartite quantum state $\psi$,
\begin{equation}
\Upsilon(A;B)_{\psi} \coloneqq \inf\left\{r : r \text{ is the achievable entanglement rate of the QSE task for } \psi \right\}.
\end{equation}
The QUI represents the information that cannot be state-exchanged using only LOCC.
Therefore, it is reasonable to quantify it by the amount of net entanglement consumed by Alice and Bob.
Before describing the main results, we review the previous bounds on the QUI.

\begin{Prop}[Previous bounds~\cite{Oppenheim}] \label{prop:Previous}
Let $\psi$ denote a pure quantum state representing quantum systems $ABR$.
The QUI $\Upsilon$ of $\psi$ is bounded as follows:
\begin{equation}
l_1[\psi] \le l_2[\psi] \le \Upsilon(A;B)_\psi \le u_1[\psi].
\end{equation}
The converse bounds $l_1$ and $l_2$, and the achievable bound $u_1$ are given by
\begin{eqnarray}
l_1[\psi]&\coloneqq& \left| S(B)_\psi- S(A)_\psi \right|, \\
l_2[\psi]&\coloneqq& \sup \left\{ S(BR_1)_{V\ket{\psi}}- S(AR_1)_{V\ket{\psi}} \right\}, \\
u_1[\psi]&\coloneqq& S(AB)_{\psi},
\end{eqnarray}
where the supremum is taken over all isometries $V_{R \to R_1R_2}$.
\end{Prop}

The quantum entropies of systems $A$ and $B$ represent the converse bound $l_1$ as their difference.
This means that when the QSE task is completed, the quantum entropies of the systems $A$ and $B$ are exchanged.
Consequently, as the entropy of one system decreases or increases, the entropy of the other system correspondingly increases or decreases.
It follows that Alice and Bob need to consume at least as much entanglement as $l_1$ to adjust their entropies.
By generalizing this notion, the authors of Ref.~\cite{Oppenheim} derived the converse bound $l_2$.
They proposed a referee-assisted scenario where a referee assists Alice and Bob by applying an isometry $V_{R \to R_1R_2}$ to the reference system $R$.
The isometry is used to distribute the referee's information to Alice and Bob without compromising the quantum entropies of systems $A$ and $B$.
Lastly, the achievability of the upper bound $u_1$ arises from the merge-and-send strategy~\cite{Oppenheim}, which consists of quantum state merging~\cite{Horodecki2005, Horodecki2007} and Schumacher compression~\cite{Schumacher1995} using ebits instead of qubit channels.

We explained the definitions of the necessary concepts for this work, including the QUI, and reviewed the previously known bounds on the QUI.
In the next section, we introduce the notion of common subspaces and provide its definition.

\section{Common subspace} \label{sec:CommonSubspace}
The merge-and-send strategy is the only known QSE protocol~\cite{Oppenheim}.
In this strategy, Alice transmits her part to Bob, and then Bob transmits all of his part to Alice.
This process involves exchanging even the QCI, which results in unnecessary entanglement consumption.
To address this issue, we investigate and formulate the concept of a common subspace that includes quantum states which do not need to be exchanged.

Let us observe the case of bipartite pure states with Schmidt decompositions~\cite{Wilde2013} given by
\begin{equation}
\ket{\phi}_{AB}=\sum_{i}\sqrt{\lambda_i}\ket{\alpha_i}_A \otimes\ket{\beta_i}_B,
\end{equation}
where Alice and Bob hold quantum systems $A$ and $B$, respectively.
The decomposition reveals correlations between $A$ and $B$; i.e., $\alpha_i$ only connects to $\beta_i$ for each $i$.
The correlations allow Alice and Bob to perform the following transformations:
\begin{equation}
\ket{\phi}_{AB}
\xrightleftharpoons[\quad V_A^\dagger \quad]{\quad V_A \quad} \ket{\phi^\mathrm{int}}_{AB}
\xrightleftharpoons[\quad W_B^\dagger \quad]{\quad W_B \quad} \ket{\phi_\mathrm{f}}_{AB}.
\end{equation}
The unitaries $V_A$ and $W_B$ satisfy $V_A\ket{\alpha_i} = \ket{\beta_i}$ and $W_B\ket{\beta_i} = \ket{\alpha_i}$ for each $i$.
We denote the state obtained by exchanging the parts $A$ and $B$ of the state $\phi$ by $\phi_\mathrm{f}$.
The intermediate state, denoted by $\phi^\mathrm{int}$, has the property $\phi^\mathrm{int} = \phi_\mathrm{f}^\mathrm{int}$, which means no local actions and entanglement consumption are required to state-exchange the state $\phi$.
In this case, local unitaries can state-exchange the initial state if and only if it can be transformed into the intermediate state by local unitaries.
For bipartite mixed states, it is challenging to transform all eigenstates into such states through fixed local unitaries.
Instead, we seek subspaces that transform only some of them.

\begin{Def}[Common subspace] \label{def:Common}
Let $\psi$ be an initial state of the QSE task representing quantum systems $ABR$.
Let $C$ denote a non-empty subspace of $\mathcal{H}_A$.
The subspace $C$ is said to be \emph{common} with respect to the initial state $\psi$ if there exist \emph{common} unitaries $V_A$ and $W_B$ that decompose $\psi$ into a \emph{common} state $\psi^\mathrm{com}$ and an \emph{uncommon} state $\psi^\mathrm{unc}$ such that
\begin{eqnarray}
\left( V_A \otimes W_B \otimes \mathds{1}_{R} \right) \ket{\psi}_{ABR}
&=& \ket{\psi^\mathrm{com}}_{ABR} + \ket{\psi^\mathrm{unc}}_{ABR}, \label{eq:Condition2} \\
\ket{\psi^\mathrm{com}}_{ABR}
&=& \ket{\psi^\mathrm{com}_\mathrm{f}}_{ABR}. \label{eq:Condition1}
\end{eqnarray}
These states are defined as follows:
\begin{eqnarray}
\ket{\psi^\mathrm{com}}_{ABR}
&\coloneqq&
\left( \Pi_{C_\mathrm{Alice}} \otimes\Pi_{C_\mathrm{Bob}} \otimes\mathds{1}_{R} \right)
\left( V_A \otimes W_B \otimes \mathds{1}_{R} \right) \ket{\psi}_{ABR}, \label{eq:COMMONstate} \\
\ket{\psi^\mathrm{unc}}_{ABR}
&\coloneqq&
\left( \Pi_{C_\mathrm{Alice}^\perp} \otimes\Pi_{C_\mathrm{Bob}^\perp} \otimes\mathds{1}_{R} \right)
\left( V_A \otimes W_B \otimes \mathds{1}_{R} \right) \ket{\psi}_{ABR}, \label{eq:UNCOMMONstate} \\
\ket{\psi^\mathrm{com}_\mathrm{f}}_{ABR}
&\coloneqq& \left( \sum_{i,j} \ket{j}\bra{i}_A \otimes \ket{i}\bra{j}_B \otimes \mathds{1}_{R} \right) \ket{\psi^\mathrm{com}}_{ABR},
\end{eqnarray}
where $C_\mathrm{Alice}$ and $C_\mathrm{Bob}$ denote the common subspaces held by Alice and Bob, respectively, i.e., $C_\mathrm{Alice} = C = C_\mathrm{Bob}$, $\Pi_C$ indicates a projection matrix on $C$, and $C^\perp$ is the orthogonal complement of $C$.
\end{Def}

\begin{figure}[t]
\includegraphics[clip,width=.85\columnwidth]{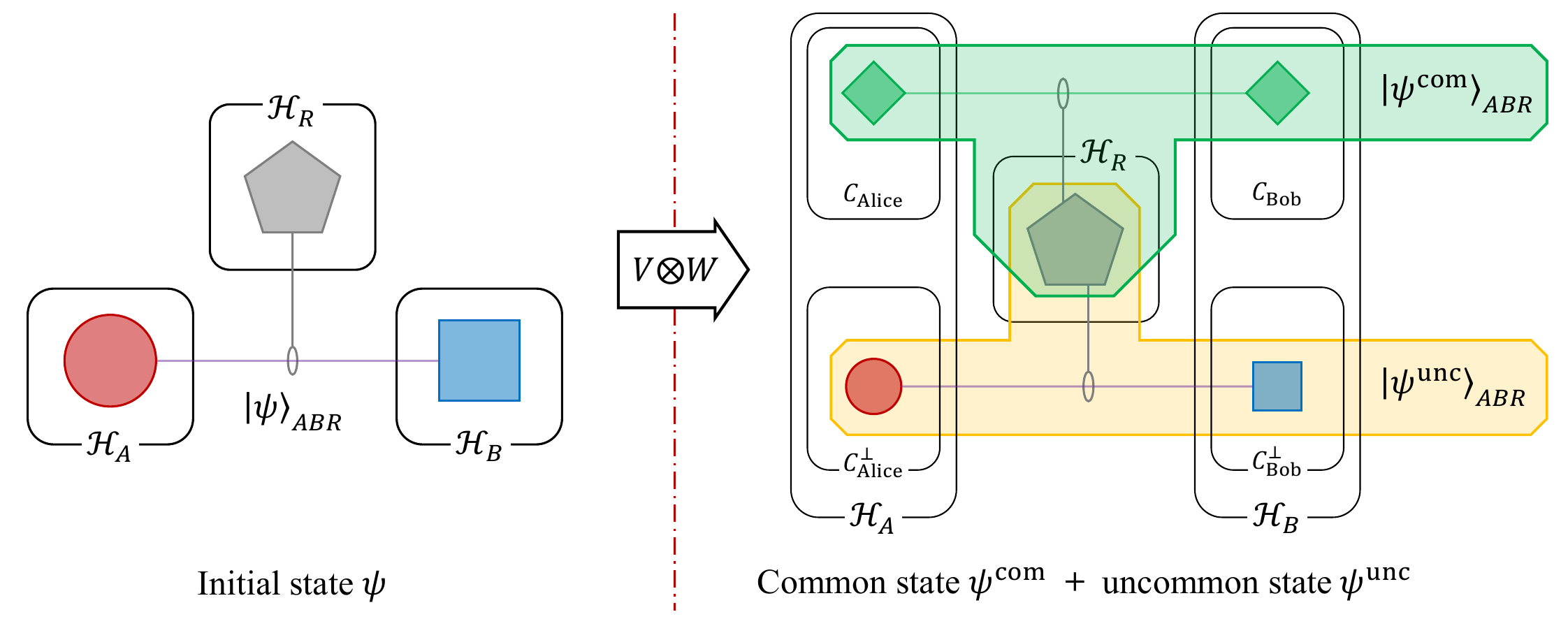}
\caption{
Illustration of common subspaces:
The Hilbert space $\mathcal{H}_A$ ($\mathcal{H}_B$) can be decomposed into a common subspace $C_\mathrm{Alice}$ ($C_\mathrm{Bob}$) and its orthogonal complement $C_\mathrm{Alice}^\perp$ ($C_\mathrm{Bob}^\perp$).
Through the common unitaries $V$ and $W$ in Definition~\ref{def:Common}, the initial state $\psi$ is transformed into the sum of a common state $\psi^\mathrm{com}$ and an uncommon state $\psi^\mathrm{unc}$, as shown in Eq.~(\ref{eq:Condition2}).
The common and uncommon states are obtained by applying $V \otimes W$ and projecting onto the subspaces $C_\mathrm{Alice} \otimes C_\mathrm{Bob} \otimes \mathcal{H}_{R}$ and $C_\mathrm{Alice}^\perp \otimes C_\mathrm{Bob}^\perp \otimes \mathcal{H}_{R}$, respectively, as shown in Eqs.~(\ref{eq:COMMONstate}) and~(\ref{eq:UNCOMMONstate}).
The common state is represented as rhombuses, and the uncommon state as a circle and a square.
By considering the common subspace, we can filter out the parts unnecessary for the QSE task, i.e., the common state, thereby reducing the amount of shared entanglement.
}
\label{fig:CSS}
\end{figure}

In Definition~\ref{def:Common}, the common unitaries reveal hidden common and uncommon states of an initial state $\psi$ while preserving its entanglement, as illustrated in Fig.~\ref{fig:CSS}.
According to the condition of Eq.~(\ref{eq:Condition1}), the common state does not need to be exchanged.
This indicates that the common state has only the QCI, while the remaining part, the uncommon state, may or may not have the QCI.
Since the common and uncommon states simultaneously represent the systems $AB$, Alice and Bob need the condition of Eq.~(\ref{eq:Condition2}) to separate and state-exchange the uncommon state.

We present the Greenberger-Horne-Zeilinger (GHZ) and Werner states~\cite{Greenberger1989, Werner1989} as simple examples.
Due to their symmetric structures, they represent common states, and their Hilbert spaces become common subspaces.
Consequently, they have only the QCI without any QUI.
On the other hand, let us consider a product state
\begin{equation}
\ket{\psi}_{ABR} = \ket{\phi^\mathrm{l}}_{AR_1} \otimes \ket{\phi^\mathrm{r}}_{R_2B},
\end{equation}
where the reference system $R$ is defined as $R = R_1 \otimes R_2$, and $\phi^\mathrm{l}$ and $\phi^\mathrm{r}$ are any bipartite states on the quantum systems $AR_1$ and $R_2B$, respectively.
Since the states $\phi^\mathrm{l}$ and $\phi^\mathrm{r}$ lack correlations, the product state $\psi$ cannot have any common subspaces.
In this case, it has only the QUI, with no QCI.

We defined the common subspace for tripartite pure states and provided examples of cases that have and do not have this subspace.
In the next section, we introduce a QSE protocol that utilizes the common subspace.

\section{Subspace exchange strategy: Achievable bound} \label{sec:SES}
Based on the notion of the common subspace, we devise a strategy for the QSE task called \emph{subspace exchange} (SSE).
This approach aims to state-exchange an uncommon state while leaving a common state intact.
It provides an achievable bound on the QUI.

\begin{figure}[t]
\hfill
\subfloat[]{\includegraphics[width=.24\linewidth,trim=0cm 0cm 0cm 0cm]{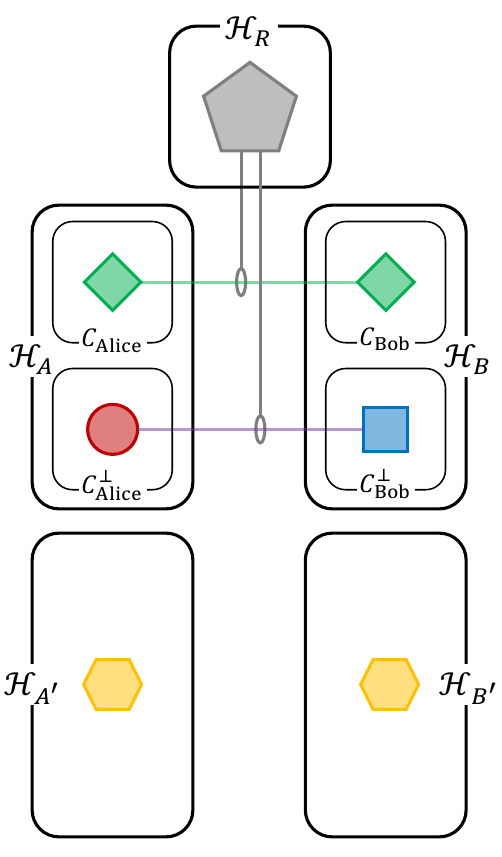}}
\hfill
\subfloat[]{\includegraphics[width=.24\linewidth,trim=0cm 0cm 0cm 0cm]{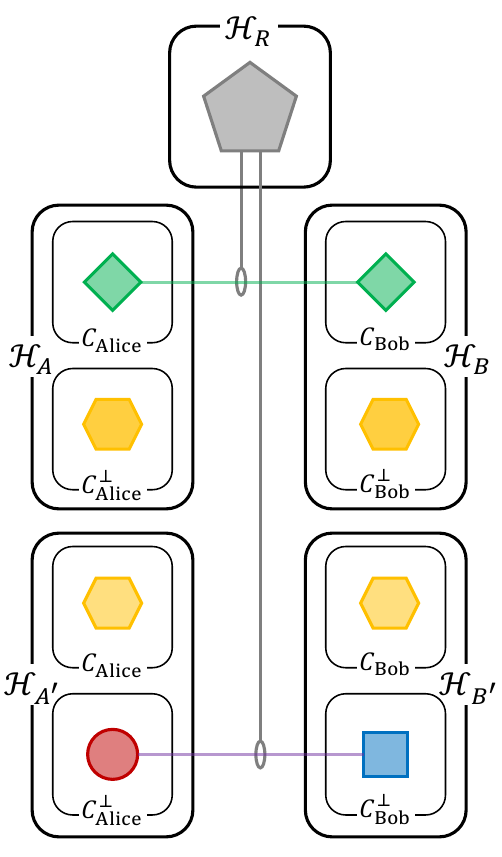}}
\hfill
\subfloat[]{\includegraphics[width=.24\linewidth,trim=0cm 0cm 0cm 0cm]{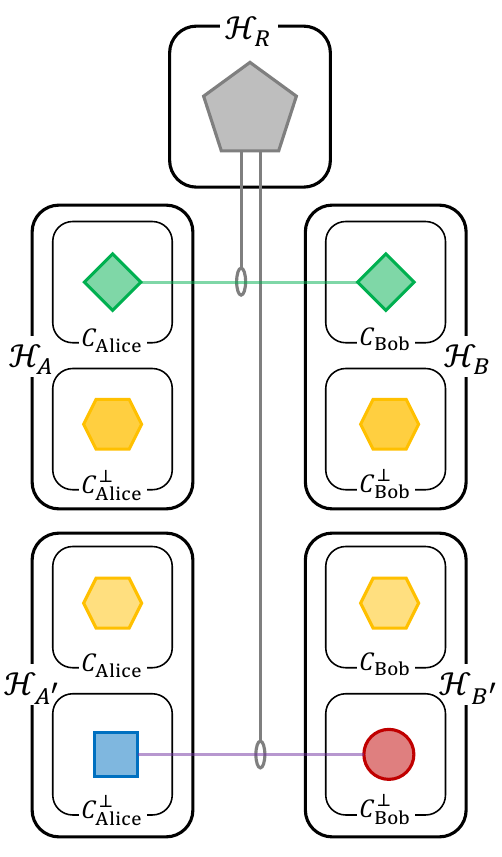}}
\hfill
\subfloat[]{\includegraphics[width=.24\linewidth,trim=0cm 0cm 0cm 0cm]{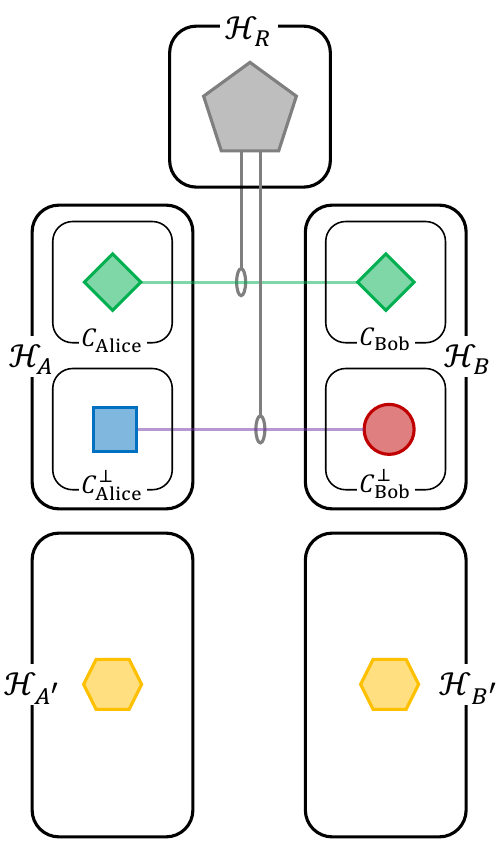}}
\hfill
\caption{
Illustration of the SSE strategy:
Alice and Bob hold quantum systems $AA'$ and $BB'$, respectively.
The SSE strategy is based on a common subspace of Definition~\ref{def:Common}.
(a) As shown in Eq.~(\ref{eq:Divided}), Alice and Bob decompose the common and uncommon states without compromising the information of the initial state.
To extract and exchange the circle and square, they need to consider the additional quantum systems $A'$ and $B'$, and prepare any pure states $\zeta$.
These additional pure states are represented as hexagons.
(b) Alice and Bob transfer the uncommon state of the systems $AB$ to the other systems $A'B'$ as in Eq.~(\ref{eq:Stretched}).
(c) They state-exchange the uncommon state of the systems $A'B'$ as in Eq.~(\ref{eq:StretchedExchanged}).
We illustrate this as the exchange of positions of the circle and square.
For convenience, the entanglement consumed or gained in this state exchange is not visualized.
(d) After the exchange, the circle and square return to the systems $AB$ as illustrated in Eq.~(\ref{eq:DividedFinal}).
}
\label{fig:SES}
\end{figure}

For an initial state $\psi$, the SSE strategy consists of five steps, as shown in Fig.~\ref{fig:SES}:
\begin{equation}
\psi
\xrightarrow[V \otimes W]{\quad\mathrm{(i)}\quad} \left( \psi^\mathrm{com} + \psi^\mathrm{unc} \right)
\xrightarrow[U \otimes U]{\quad\mathrm{(ii)}\quad} \psi^\mathrm{str}
\xrightarrow[\quad\text{Exchanging uncommon state}\quad]{\quad\mathrm{(iii)}\quad} \psi^\mathrm{str}_\mathrm{f}
\xrightarrow[U^\dagger \otimes U^\dagger]{\quad\mathrm{(iv)}\quad} \left( \psi^\mathrm{com} + \psi^\mathrm{unc}_\mathrm{f} \right)
\xrightarrow[W^\dagger \otimes V^\dagger]{\quad\mathrm{(v)}\quad} \psi_\mathrm{f}.
\end{equation}
(i) First, for a common subspace $C$, Alice and Bob apply the corresponding common unitaries $V_A$ and $W_B$ from Definition~\ref{def:Common} to decompose a common state $\psi^\mathrm{com}$ and an uncommon state $\psi^\mathrm{unc}$, as shown in Fig.~\ref{fig:SES}(a):
\begin{equation} \label{eq:Divided}
\ket{\psi}_{ABR}
\xrightarrow{V_A \otimes W_B}
\ket{\psi^\mathrm{com}}_{ABR} + \ket{\psi^\mathrm{unc}}_{ABR}.
\end{equation}
Since the common state has already been state-exchanged, i.e., $\psi^\mathrm{com} = \psi^\mathrm{com}_\mathrm{f}$, only the uncommon state $\psi^\mathrm{unc}$ needs to be state-exchanged.
(ii) The problem is that the common and uncommon states represent the same quantum systems $ABR$.
To separate them, Alice and Bob consider additional systems $A'$ and $B'$ whose dimensions are equal to those of systems $A$ and $B$, respectively.
They apply local unitaries $U_{AA'}$ and $U_{BB'}$ to transfer their parts of the uncommon state to the systems $A'$ and $B'$ as follows:
\begin{equation} \label{eq:Stretched}
\left( \ket{\psi^\mathrm{com}}_{ABR} + \ket{\psi^\mathrm{unc}}_{ABR} \right) \otimes \ket{\zeta}_{A'} \otimes \ket{\zeta}_{B'}
\xrightarrow{U_{AA'} \otimes U_{BB'}} \ket{\psi^\mathrm{str}}_{ABRA'B'}
\coloneqq \ket{\psi^\mathrm{com}}_{ABR} \otimes \ket{\zeta}_{A'} \otimes \ket{\zeta}_{B'} + \ket{\eta}_{A} \otimes \ket{\eta}_{B} \otimes \ket{\psi^\mathrm{unc}}_{RA'B'},
\end{equation}
where $\zeta$ and $\eta$ denote any pure states in the common subspace $C$ and its orthogonal complement $C^\perp$, respectively.
The choice of $\zeta$ and $\eta$ does not affect the amount of entanglement consumed in this strategy.
The unitaries $U_{AA'}$ and $U_{BB'}$ are constructed based on the common subspace $C$.
Specifically, given the basis vectors spanning the common subspace, $U_{AA'}$ and $U_{BB'}$ can be described.
We present a detailed explanation of this construction in Appendix~\ref{app:SES}.
We refer to the resulting state as a \emph{stretched} state $\psi^\mathrm{str}$ for the initial state $\psi$, illustrated in Fig.~\ref{fig:SES}(b).
(iii) Alice and Bob exchange the parts $A'$ and $B'$ of the stretched state $\psi^\mathrm{str}$, as depicted in Fig.~\ref{fig:SES}(c).
To do this, they use the merge-and-merge strategy~\cite{Lee2019a}, which consists of two quantum state merging protocols~\cite{Horodecki2005, Horodecki2007}.
In the first merging, Alice state-merges her part $A'$ to Bob, and in the second merging, Bob state-merges his part $B'$ to Alice.
The quantum conditional entropies $S(A'|BB')_{\psi^\mathrm{str}}$ and $S(B'|A)_{\psi^\mathrm{str}}$ denote the optimal achievable entanglement rates for the two mergings:
\begin{equation} \label{eq:StretchedExchanged}
\ket{\psi^\mathrm{str}}_{ABRA'B'}
\xrightarrow{\text{merge-and-merge}} \ket{\psi^\mathrm{str}_\mathrm{f}}_{ABRA'B'}
\coloneqq \ket{\psi^\mathrm{com}}_{ABR} \otimes\ket{\zeta}_{A'} \otimes\ket{\zeta}_{B'} + \ket{\eta}_{A} \otimes\ket{\eta}_{B} \otimes\ket{\psi^\mathrm{unc}_\mathrm{f}}_{RA'B'},
\end{equation}
where $\psi^\mathrm{str}_\mathrm{f}$ denotes the exchanged state obtained by exchanging the parts $A'$ and $B'$ of the stretched state $\psi^\mathrm{str}$.
(iv) Alice and Bob make the common and uncommon states represent the same quantum systems $ABR$ again, as depicted in Fig.~\ref{fig:SES}(d):
\begin{equation} \label{eq:DividedFinal}
\ket{\psi^\mathrm{str}_\mathrm{f}}_{ABRA'B'}
\xrightarrow{U^\dagger_{AA'} \otimes U^\dagger_{BB'}} \left( \ket{\psi^\mathrm{com}}_{ABR} + \ket{\psi^\mathrm{unc}_\mathrm{f}}_{ABR} \right) \otimes\ket{\zeta}_{A'} \otimes\ket{\zeta}_{B'},
\end{equation}
where $U^\dagger$ denotes the inverse of the unitary $U$ in Eq.~(\ref{eq:Stretched}).
(v) Finally, Alice and Bob apply the inverses of the common unitaries $V_A$ and $W_B$ in Eq.~(\ref{eq:Divided}) to obtain the final state $\psi_\mathrm{f}$:
\begin{equation}
\ket{\psi^\mathrm{com}}_{ABR} + \ket{\psi^\mathrm{unc}_\mathrm{f}}_{ABR}
\xrightarrow{W_A^\dagger \otimes V_B^\dagger} \ket{\psi_\mathrm{f}}_{ABR}.
\end{equation}

Consequently, the SSE strategy accomplishes the QSE task.
Detailed explanations of the related states and the LOCC for each step are presented in Appendix~\ref{app:SES}.
The achievable entanglement rate for the SSE strategy provides an achievable bound on the QUI.

\begin{Thm}[Achievable bound] \label{thm:Achievable}
Let $\psi$ be a pure state representing quantum systems $ABR$.
The QUI $\Upsilon$ of $\psi$ is upper-bounded by $u_\mathrm{new}$ as follows:
\begin{equation}
\Upsilon(A;B)_\psi \le u_\mathrm{new}[\psi^\mathrm{str}] \coloneqq S(R|A)_{\psi^\mathrm{str}},
\end{equation}
where the stretched state $\psi^\mathrm{str}$ for $\psi$ is defined in Eq.~(\ref{eq:Stretched}).
\end{Thm}

The merge-and-merge strategy is the only part consuming shared entanglement in the SSE strategy.
Thus, the sum of the achievable entanglement rates for the two mergings becomes an achievable entanglement rate of the SSE strategy:
\begin{equation}
S(A'|BB')_{\psi^\mathrm{str}} + S(B'|A)_{\psi^\mathrm{str}} = u_\mathrm{new}[\psi^\mathrm{str}].
\end{equation}
Even if Bob initiates the merge-and-merge strategy by first merging his part $B'$, the achievable entanglement rate of the SSE strategy still remains $S(R|A)$.

In this section, we introduced the SSE strategy using common subspaces of the initial state.
If the initial state has no common subspaces, then the SSE strategy reduces to the merge-and-send strategy.
We prove the achievability of the upper bound $u_\mathrm{new}$ in Appendix~\ref{app:Achievability}.
Additionally, we show in Sec.~\ref{sec:Tightness} that the bound $u_\mathrm{new}$ requires less entanglement compared to the previous bound $u_1$.
In the next section, we introduce a generalized QSE task to find a converse bound on the QUI.

\section{referee-assisted exchange task: Converse bound} \label{sec:RAE}
A converse bound on the QUI provides a theoretical limit on entanglement consumption for the QSE task.
Alice and Bob must spend at least this amount of entanglement to achieve their task successfully.
To find a converse bound, we devise a \emph{referee-assisted exchange} (RAE) task, which includes the QSE task as a special case.

\begin{figure}[t]
\hfill
\subfloat[]{\includegraphics[width=.42\linewidth,trim=0cm 0cm 0cm 0cm]{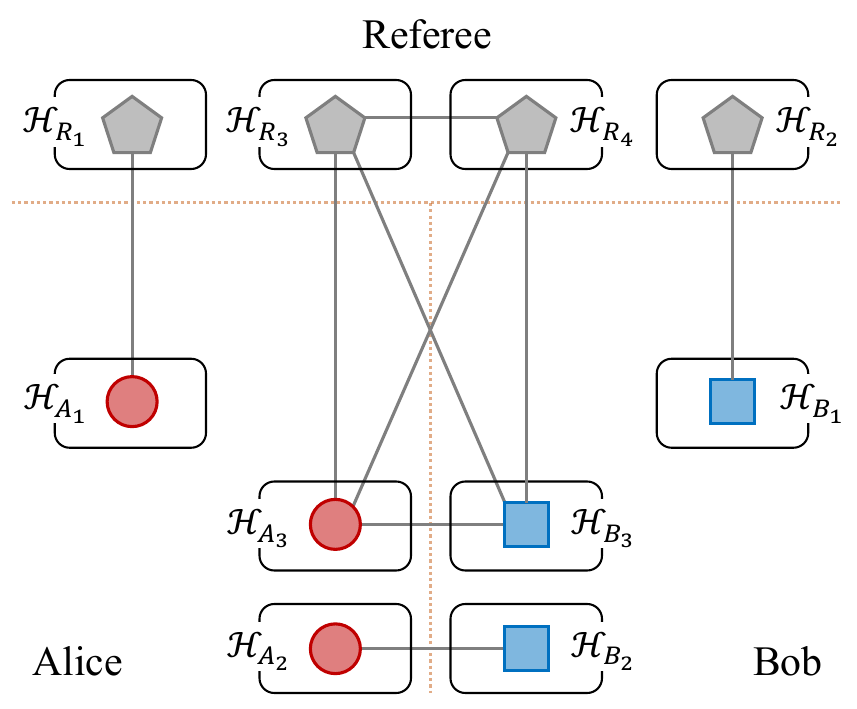}}
\quad\quad
\subfloat[]{\includegraphics[width=.42\linewidth,trim=0cm 0cm 0cm 0cm]{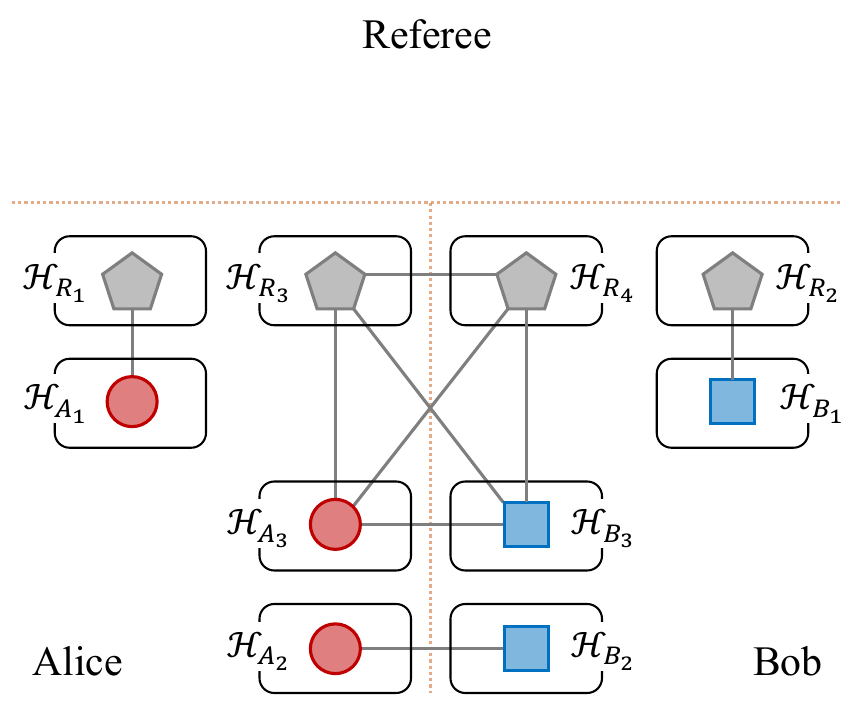}}
\hfill
\bigskip
\hfill
\subfloat[]{\includegraphics[width=.42\linewidth,trim=0cm 0cm 0cm 0cm]{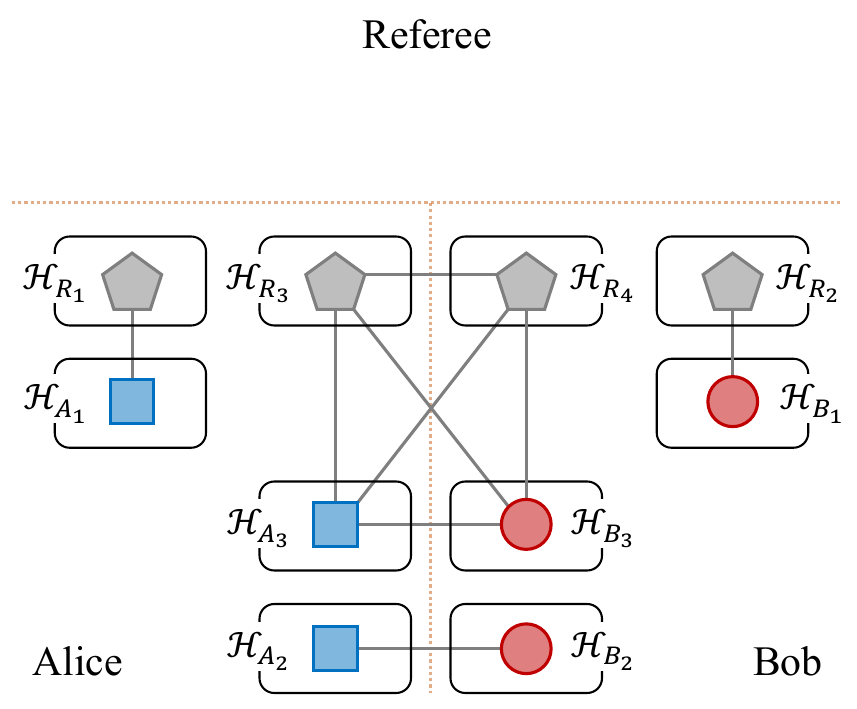}}
\quad\quad
\subfloat[]{\includegraphics[width=.42\linewidth,trim=0cm 0cm 0cm 0cm]{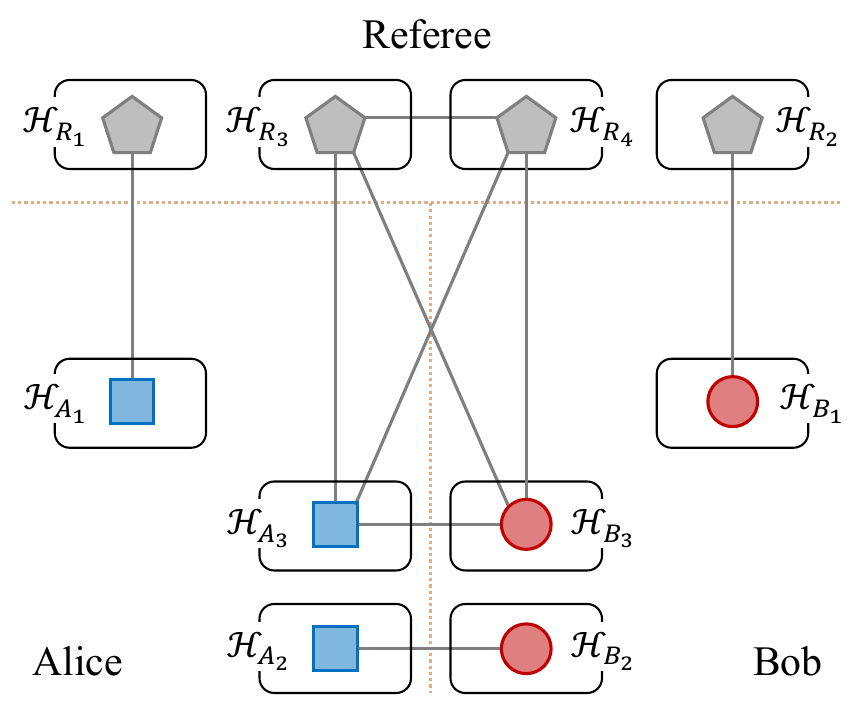}}
\hfill
\caption{
Steps of the RAE task:
With the assistance of the referee, Alice and Bob can perform the QSE task more efficiently as outlined in Eq.~(\ref{eq:RAE}).
(a) First, through three-party LOCC of Eq.~(\ref{eq:Reversible}), Alice, Bob, and the referee decompose the initial state $\psi$ into the decomposed state $\psi^\mathrm{dec}_n$, which consists of four pure quantum states.
(b) In the second step, the referee distributes his quantum states to Alice and Bob.
We denote the quantum systems of the quantum states transmitted to Alice and Bob with the same labels as those of the referee's quantum system.
(c) Thirdly, Alice and Bob exchange their parts $A_i$ and $B_i$ with the referee's assistance.
The resulting state is presented in Eq.~(\ref{eq:DecomposedFinal}).
(d) Alice and Bob return the quantum states received from the referee.
}
\label{fig:RAE}
\end{figure}

The RAE task is a three-party scenario involving Alice, Bob, and a referee, where the referee assists Alice and Bob in performing the QSE task for an initial state $\psi$.
The initial state $\psi$ represents quantum systems $ABR$, where Alice, Bob, and the referee share quantum systems $A$, $B$, and $R$, respectively.
The final state $\psi_\mathrm{f}$ for the RAE task is defined as shown in Eq.~(\ref{eq:Final}).
In the RAE task, the referee provides two types of assistance:
\begin{enumerate}
\item \textbf{Three-party LOCC:} The referee can perform reversible transformations in collaboration with Alice and Bob. Such transformations $\Lambda$ decompose $n$ copies of the initial state into four pure states with non-negative rates $r_i$:
\begin{equation} \label{eq:Reversible}
\ket{\psi}_{ABR}^{\otimes n}
\xrightleftharpoons[\quad \Lambda_n^{- 1} \quad]{\quad \Lambda_n \quad}
\ket{\psi^\mathrm{dec}_n}
\coloneqq
\ket{\phi^\mathrm{l}}_{A_1R_1}^{ \otimes \lfloor r_1 n \rfloor} \otimes
\ket{\phi^\mathrm{b}}_{A_2B_2}^{ \otimes \lfloor r_2 n \rfloor} \otimes
\ket{\phi^\mathrm{r}}_{R_2B_1}^{ \otimes \lfloor r_3 n \rfloor} \otimes
\ket{\phi^\mathrm{c}}_{A_3R_3R_4B_3}^{ \otimes \lfloor r_4 n \rfloor},
\end{equation}
where Alice, Bob, and the referee hold quantum systems $A_i$, $B_i$, and $R_i$, respectively.
They share four pure states $\phi^\mathrm{l}$, $\phi^\mathrm{b}$, $\phi^\mathrm{r}$, and $\phi^\mathrm{c}$, as depicted in Fig.~\ref{fig:RAE}(a).
The errors of the transformations converge to zero as $n$ approaches infinity.
The output state, $\psi^\mathrm{dec}_n$, of the protocol $\Lambda_n$ is referred to as a \emph{decomposed} state. 

\item \textbf{Information transmission and entanglement counting:}
The referee can send his quantum states to Alice or Bob and receive them back at any time.
The referee can share unlimited entanglement with Alice or Bob for information transmission.
Since the QUI is quantified by the entanglement consumed by Alice and Bob, the entanglement used in this transmission is not considered.
In addition, the referee can only transfer the states that he initially possessed and cannot receive the states sent by Alice and Bob.
Without these assumptions, Alice and Bob could complete the QSE task without spending any of their entanglement.
\end{enumerate}

For each $n$, an RAE protocol $\mathcal{E}_n^{\mathrm{RAE}}$ for $\psi^{\otimes n}$ comprises five steps, as shown in Fig.~\ref{fig:RAE}:
\begin{equation} \label{eq:RAE}
\psi^{\otimes n}
\xrightarrow[\quad\text{3-party }\Lambda_n\quad]{\quad\mathrm{(i)}\quad} \psi^\mathrm{dec}_n
\xrightarrow[\quad\text{Referee leaves}\quad]{\quad\mathrm{(ii)}\quad} \psi^\mathrm{dec}_n
\xrightarrow[\quad\text{3-party }\mathcal{\bar{E}}_n\quad]{\quad\mathrm{(iii)}\quad} \psi^\mathrm{dec}_{n,\mathrm{f}}
\xrightarrow[\quad\text{Referee comes back in }\quad]{\quad\mathrm{(iv)}\quad} \psi^\mathrm{dec}_{n,\mathrm{f}}
\xrightarrow[\quad\text{3-party }\Lambda_{n,\mathrm{f}}^{- 1}\quad]{\quad\mathrm{(v)}\quad} \psi_\mathrm{f}^{\otimes n}.
\end{equation}
(i) Alice, Bob, and the referee apply the transformation $\Lambda_n$ in Eq.~(\ref{eq:Reversible}) to $\psi^{\otimes n}$ to obtain the decomposed state $\psi^\mathrm{dec}_n$, as illustrated in Fig.~\ref{fig:RAE}(a).
From the perspective of the QSE task, Alice and Bob can ignore the bottom state $\phi^\mathrm{b}$ since it can be state-exchanged without shared entanglement.
In addition, the left and right states $\phi^\mathrm{l}$ and $\phi^\mathrm{r}$ can be optimally state-exchanged using Schumacher compression~\cite{Schumacher1995}.
Thus, they can focus on the center state $\phi^\mathrm{c}$ instead of the initial state $\psi$.
(ii) The referee transmits his parts $R_1R_3$ and $R_2R_4$ to Alice and Bob, respectively, as depicted in Fig.~\ref{fig:RAE}(b).
(iii) Alice and Bob, with the referee's assistance, exchange their parts $A_i$ and $B_i$ of the decomposed state $\psi^\mathrm{dec}_n$.
This is represented by a three-party LOCC $\mathcal{\bar{E}}_n$.
Note that the referee can assist the others during the protocol $\mathcal{\bar{E}}_n$.
While any pair of the three parties can share and use entanglement, only the entanglement between Alice and Bob is counted.
After applying LOCC $\mathcal{\bar{E}}_n$, Alice and Bob share an exchanged state, as shown in Fig.~\ref{fig:RAE}(c),
\begin{equation} \label{eq:DecomposedFinal}
\ket{\psi_{n,\mathrm{f}}^\mathrm{dec}}\coloneqq
\ket{\phi^\mathrm{l}_\mathrm{f}}_{B_1R_1}^{ \otimes \lfloor r_1 n \rfloor} \otimes
\ket{\phi^\mathrm{b}_\mathrm{f}}_{A_2B_2}^{ \otimes \lfloor r_2 n \rfloor} \otimes
\ket{\phi^\mathrm{r}_\mathrm{f}}_{R_2A_1}^{ \otimes \lfloor r_3 n \rfloor} \otimes
\ket{\phi^\mathrm{c}_\mathrm{f}}_{A_3R_3R_4B_3}^{ \otimes \lfloor r_4 n \rfloor},
\end{equation}
where $\psi_{n,\mathrm{f}}^\mathrm{dec}$ denotes the state after the exchange.
(iv) Alice and Bob return the parts $R_1R_3$ and $R_2R_4$ of $\psi_{n,\mathrm{f}}^\mathrm{dec}$ to the referee, as shown in Fig.~\ref{fig:RAE}(d).
(v) Lastly, Alice and Bob swap and play their respective roles in LOCC $\Lambda_n^{- 1}$ to obtain $n$ copies of the final state $\psi_\mathrm{f}$.
This modified LOCC is denoted by $\Lambda_{n,\mathrm{f}}^{- 1}$.

For sufficiently large $n$, the errors of the RAE protocols vanish, and the RAE task completes successfully.
Remark that with the referee's assistance, Alice and Bob can employ more protocols, including all QSE protocols operating on the initial state $\psi$.
In the third stage of the RAE task, they can convert any QSE protocol $\mathcal{E}_n$ in Eq.~(\ref{eq:TraceNorm}) to the three-party protocol $\mathcal{\bar{E}}_n$ in Eq.~(\ref{eq:RAE}).
This allows Alice and Bob to exchange their parts $A$ and $B$ more efficiently, consuming less entanglement.
Thus, any achievable entanglement rate of the QSE task becomes that of the RAE task.
This provides a converse bound on the QUI.

\begin{Thm}[Converse bound] \label{thm:Converse}
Let $\psi$ be a pure state representing quantum systems $ABR$.
The QUI $\Upsilon$ of $\psi$ is lower-bounded by $l_\mathrm{new}$ as follows:
\begin{equation} \label{eq:thm:Converse}
l_\mathrm{new}\left[\Lambda\right]
\coloneqq 
r_1 S(A_1)_{\phi^\mathrm{l}}
+ r_3 S(B_1)_{\phi^\mathrm{r}}
+ r_4 \left( S(B_3R_3)_{\phi^\mathrm{c}} - S(A_3R_3)_{\phi^\mathrm{c}} \right)
\le \Upsilon(A;B)_\psi,
\end{equation}
where the states $\phi^\mathrm{l}$, $\phi^\mathrm{r}$, and $\phi^\mathrm{c}$ are presented in the reversible transformation $\Lambda$ of Eq.~(\ref{eq:Reversible}).
Moreover, the supremum taken over all such transformations provides the tighter bound on the QUI:
\begin{equation}
l_\mathrm{new}[\psi] \coloneqq \sup l_\mathrm{new}\left[\Lambda\right] \le \Upsilon(A;B)_\psi.
\end{equation}
\end{Thm}

We introduced the RAE task, where a third-party referee assists Alice and Bob in efficiently performing the QSE task.
We provide the proof of Theorem~\ref{thm:Converse} in Appendix~\ref{app:Converse}.
In the next section, we demonstrate that our bounds are tighter than the previous bounds and validate them with examples.

\section{Tightness of our bounds and its demonstration} \label{sec:Tightness}
We show that our bounds on the QUI are tighter than the previous ones.
They provide a better estimation of the QUI.
To demonstrate this, we calculate our bounds for a non-trivial example and compare their graphs. 

The tightness of the achievable bound $u_\mathrm{new}$ arises from the relationship between an initial state $\psi$ and its stretched state $\psi^\mathrm{str}$.
In the SSE strategy, Alice and Bob's local unitaries transform the initial state into the stretched state.
Consequently, the entropy of the reference system $R$ remains unchanged, i.e.,
\begin{equation}
S(AB)_\psi = S(R)_\psi = S(R)_{\psi^\mathrm{str}} = S(ABA'B')_{\psi^\mathrm{str}}.
\end{equation}
Then, the non-negativity of quantum mutual information~\cite{Araki1970} ensures the tightness of the achievable bound $u_\mathrm{new}$, i.e.,
\begin{equation}
u_\mathrm{new}[\psi^\mathrm{str}] \le u_1[\psi].
\end{equation}
Next, we consider a special case of the RAE task where the referee's state of the reference system $R$ is split into two using an isometry $V_{R \to R_3R_4}$:
\begin{equation}
\ket{\psi}_{ABR} \xrightleftharpoons{\quad\text{Isometry }V\quad} \ket{\phi^\mathrm{c}}_{AR_3R_4B}.
\end{equation}
After the splitting, the referee transmits the parts $R_3$ and $R_4$ to Alice and Bob, respectively. Such isometries are a special case of reversible transformations $\Lambda$ of Eq.~(\ref{eq:Reversible}) since they can apply to each copy of the initial state and are reversible.
This shows the tightness of the converse bound $l_\mathrm{new}[\psi]$.
As a result, our bounds are tighter than the previous ones:
\begin{equation} \label{eq:Tightness}
l_2[\psi] \le l_\mathrm{new}[\psi] \le \Upsilon(A;B)_\psi \le u_\mathrm{new}[\psi^\mathrm{str}] \le u_1[\psi].
\end{equation}

To illustrate the usefulness of our bounds, one needs to know all reversible transformations for a given initial state.
Since this is generally difficult, we consider a class of tripartite pure states~\cite{Vidal2000}
\begin{equation} \label{eq:Zeta}
\ket{\zeta}_{ABR}
=\frac{c_0}{\sqrt{2}}\left(\ket{000}_{ABR}+ \ket{011}_{ABR}\right)
+ \frac{c_1}{\sqrt{2}}\left(\ket{122}_{ABR}+ \ket{223}_{ABR}\right)
+ \frac{c_2}{\sqrt{2}}\left(\ket{334}_{ABR}+ \ket{444}_{ABR}\right)
+ c_3\ket{555}_{ABR},
\end{equation}
where $c_i$ are non-negative numbers with $\sum_{i=0}^3 c_i^2=1$, and verify the following inequalities instead:
\begin{equation} \label{eq:weakTightness}
l_1[\zeta] \le l_\mathrm{new}[\Lambda'] \le \Upsilon(A;B)_\zeta \le u_\mathrm{new}[\zeta^\mathrm{str}] \le u_1[\zeta].
\end{equation}
Here, $\Lambda'$ denotes a reversible transformation~\cite{Vidal2000} converting the initial state $\zeta$ into a combination of Einstein-Podolsky-Rosen (EPR) and GHZ states as follows:
\begin{equation} \label{eq:AsymptoticDecomposition}
\ket{\zeta}_{ABR}^{\otimes n}
\xrightleftharpoons[\quad \Lambda'^{- 1}_n \quad]{\quad \Lambda'_n \quad}
\ket{\mathrm{EPR}}_{A_1R_1}^{ \otimes \lfloor c_1^2 n \rfloor} \otimes
\ket{\mathrm{EPR}}_{A_2B_2}^{ \otimes \lfloor c_2^2 n \rfloor} \otimes
\ket{\mathrm{EPR}}_{R_2B_1}^{ \otimes \lfloor c_0^2 n \rfloor} \otimes
\ket{\mathrm{GHZ}}_{A_3R_3B_3}^{ \otimes \left\lfloor (- \sum_{i=0}^3 c_i^2\log c_i^2) n \right\rfloor},
\end{equation}
which is a special case of the reversible transformation of Eq.~(\ref{eq:Reversible}).
To calculate the upper bound $u_\mathrm{new}[\zeta^\mathrm{str}]$, we choose a common subspace defined as follows:
\begin{equation} \label{eq:CommonSubspaceExample}
C' = \mathrm{span}\{\ket{3}, \ket{4}, \ket{5}\}.
\end{equation}
Without considering common unitaries, we can find the common and uncommon states corresponding to the common subspace $C'$ as follows:
\begin{eqnarray}
\ket{\zeta^\mathrm{com}}_{ABR}
&=& \Pi_{C'} \otimes\Pi_{C'} \otimes\mathds{1}_{R} \ket{\zeta}_{ABR}
= \frac{c_2}{\sqrt{2}}\left(\ket{334}_{ABR}+ \ket{444}_{ABR}\right)+ c_3\ket{555}_{ABR}, \\
\ket{\zeta^\mathrm{unc}}_{ABR}
&=&\Pi_{C'^\perp} \otimes\Pi_{C'^\perp} \otimes\mathds{1}_{R} \ket{\zeta}_{ABR}
= \frac{c_0}{\sqrt{2}}\left(\ket{000}_{ABR}+ \ket{011}_{ABR}\right)
+ \frac{c_1}{\sqrt{2}}\left(\ket{122}_{ABR}+ \ket{223}_{ABR}\right).
\end{eqnarray}
The corresponding stretched state $\zeta^\mathrm{str}$ is represented as
\begin{equation} \label{eq:ZetaDivided}
\ket{\zeta^\mathrm{str}}_{ABRA'B'}
=\ket{\zeta^\mathrm{com}}_{ABR} \otimes\ket{3}_{A'} \otimes\ket{3}_{B'}
+ \ket{0}_{A} \otimes\ket{0}_{B} \otimes\ket{\zeta^\mathrm{unc}}_{RA'B'}.
\end{equation}

\begin{figure}[t]
\includegraphics[clip,width=.7\columnwidth]{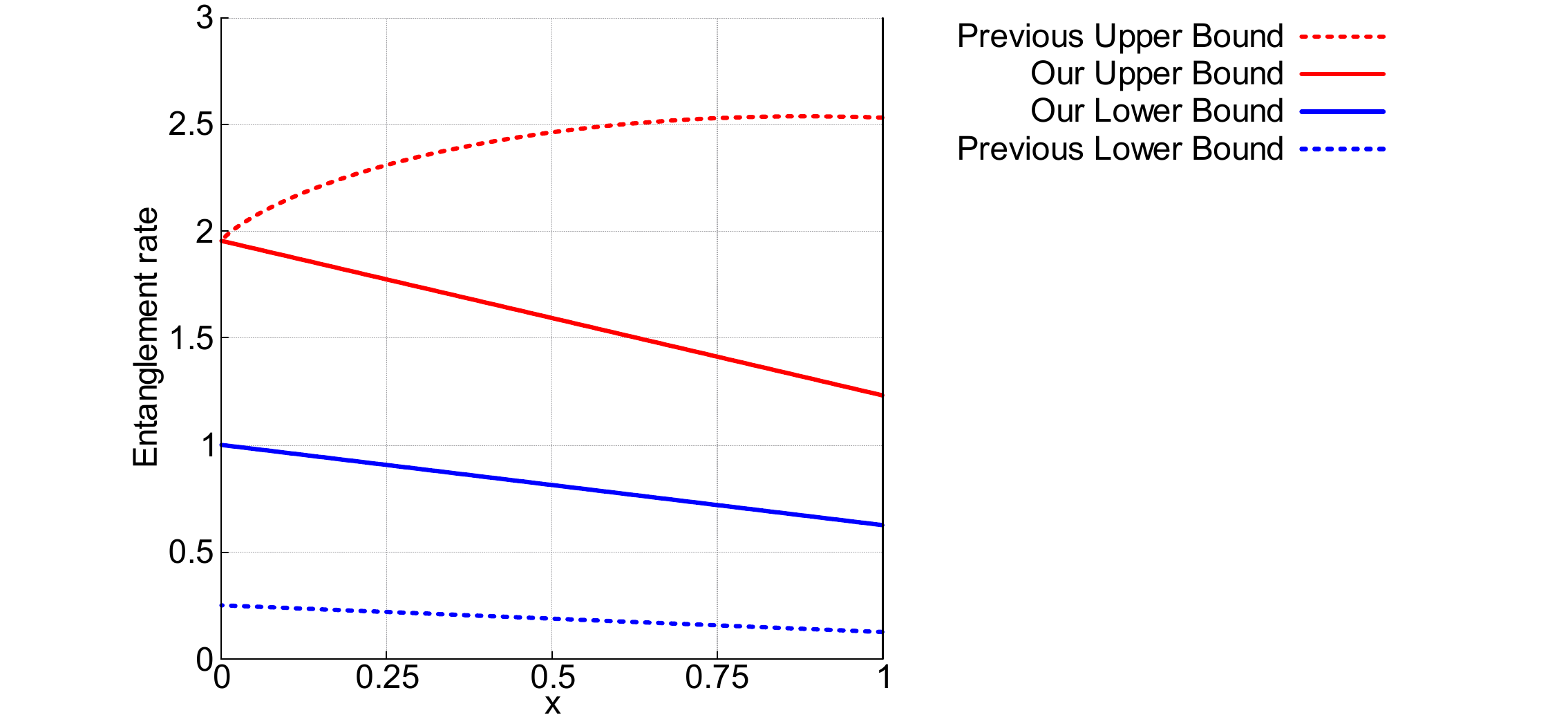}
\caption{
Comparison between our bounds on the QUI and the previous ones:
To draw graphs of the bounds in Eq.~(\ref{eq:weakTightness}), we parameterize an initial state $\zeta$ in Eq.~(\ref{eq:Zeta}) through Eq.~(\ref{eq:Parameter}).
With the parameter $x \in [0, 1]$, we express the graph of our bounds as straight lines and the previous bounds as the dotted lines.
The graphs verify that our bounds are tighter than the previous ones.
}
\label{fig:Graphs}
\end{figure}

Then, we can calculate the aforementioned bounds as
\begin{eqnarray}
l_1[\zeta]
&=& |c_0^2- c_1^2|, \\
l_\mathrm{new}[\Lambda']
&=& c_0^2+ c_1^2, \\
u_\mathrm{new}[\zeta^\mathrm{str}]
&=& - c_0^2\log \frac{c_0^2}{2}+ (c_0^2+ c_1^2)\log (c_0^2+ c_1^2)- c_1^2\log \frac{c_1^2}{2}, \\
u_1[\zeta]
&=& - c_0^2\log \frac{c_0^2}{2} - c_1^2\log \frac{c_1^2}{2} - c_2^2\log c_2^2 - c_3^2\log c_3^2.
\end{eqnarray}
To draw their graphs, we parameterize the coefficients $c_i$ as
\begin{equation} \label{eq:Parameter}
c_0=\sqrt{\frac{5- 2x}{8}},
\quad c_1=\sqrt{\frac{3-x}{8}},
\quad c_2=\sqrt{\frac{x}{8}},
\quad c_3=\sqrt{\frac{x}{4}},
\end{equation}
where $0 \le x \le 1$.
In Fig.~\ref{fig:Graphs}, we present the graphs of the bounds for the parametrized state of $\zeta$.
The graphs demonstrate that our bounds offer a more accurate estimate of the QUI compared to the previous bounds.

\section{Applications to other communication scenarios} \label{sec:Applications}
The concept of the common subspace can be applied to other quantum communication scenarios.
The first example is an exact single-shot scenario of the QSE task, where two parties transform one initial state into one final state without error.
Additionally, our concept can be extended and applied to quantum communication tasks involving three or more parties.
In particular, we consider the quantum state rotation (QSR) task~\cite{Lee2021} and propose an efficient protocol based on a three-party common subspace.
In both examples, the parties can achieve their tasks more efficiently by spending less entanglement.

\subsection{Exact single-shot scenario: Non-local SWAP}
The first application of our results is an exact single-shot version~\cite{Lee2019} of the QSE task.
In this version, Alice and Bob must exactly state-exchange a single copy of an initial state without error.
This is similar to non-local SWAP operations in distributed quantum computation~\cite{Hammerer2002, Yimsiriwattana2004, Zhang2005}.
To our knowledge, it is known that two remote parties can non-locally swap their respective qubits by consuming two shared ebits and four noiseless bit channels~\cite{Collins2001}.

We explain that using a common subspace for a target state allows for a more efficient implementation requiring less shared entanglement.
First, consider the tripartite state $\zeta$ from Eq.~(\ref{eq:Zeta}), and assume Alice and Bob have no information about the initial state $\zeta$, i.e., they do not know any common subspace of $\zeta$.
In this case, they use quantum teleportation for qudits~\cite{Bennett1993} twice to perform a non-local SWAP operation.
These teleportation protocols require two pure maximally entangled states with Schmidt ranks of six, resulting in an entanglement consumption of $2\log 6\approx 5.1699$ ebits.

Conversely, if Alice and Bob are aware of the common subspace from Eq.~(\ref{eq:CommonSubspaceExample}), they can convert their state $\zeta$ to the stretched state $\zeta^\mathrm{str}$ of Eq.~(\ref{eq:ZetaDivided}), and complete their non-local SWAP by only exchange the parts $A'$ and $B'$ of $\zeta^\mathrm{str}$.
Given the reduced dimensions of the supports of the exchanged parts, less entanglement is required for information transmission, i.e., Alice and Bob spend only four ebits to implement the non-local SWAP of $A'$ and $B'$.
Consequently, they can save up to $2\log 6- 4\approx 1.1699$ ebits for the non-local SWAP.

This example demonstrates that in the exact single-shot scenario of the QSE task, the concept of the common subspace is applicable. One can design efficient circuit implementations to non-locally swap bipartite mixed states with reduced shared entanglement.

\subsection{Three-party scenario: QSR task} \label{sec:QSR}
As the second application, we consider a QSR task~\cite{Lee2021} for three parties, where the $i$th party transmits their quantum state to the $(i+ 1)$th party.
By extending the notion of the common subspace, the parties can complete the QSR task using less entanglement than the previously known method.

Let $\xi$ denote an initial state of the QSR task, representing the quantum systems $A_1A_2A_3R$, where each $i$-th party holds a quantum system $A_i$.
The goal of the three parties is to circularly transmit their respective states using LOCC and bipartite entanglement.
The final state of the QSR task is given by
\begin{equation} \label{eq:XiFinal}
\ket{\xi_\mathrm{f}}_{A_1A_2A_3R}
\coloneqq \left( \sum_{i,j,k} \ket{k}\bra{i}_{A_1} \otimes \ket{i}\bra{j}_{A_2} \otimes \ket{j}\bra{k}_{A_3} \right) \otimes \mathds{1}_{R}
\ket{\xi}_{A_1A_2A_3R}.
\end{equation}
The sum of bipartite entanglement between any two parties defines an achievable entanglement rate for the QSR task.
The main question is to find a protocol with minimal entanglement to achieve the QSR task.

The only known strategy to perform the QSR task is the three-party merge-and-send strategy~\cite{Lee2021}, where the parties sequentially transmit their respective states using quantum state merging~\cite{Horodecki2007} and Schumacher compression~\cite{Schumacher1995}:
\begin{equation}
1^{\mathrm{st}}\text{ party}
\xrightarrow[]{\quad\text{Merging }A_1\quad} 2^{\mathrm{nd}}\text{ party}
\xrightarrow[]{\quad\text{Merging }A_2\quad} 3^{\mathrm{rd}}\text{ party}
\xrightarrow[]{\quad\text{Sending }A_3\quad} 1^{\mathrm{st}}\text{ party}.
\end{equation}
In this strategy, the first party begins by state-merging their part $A_1$ to the second party.
Depending on who starts the three-party merge-and-send strategy, different achievable entanglement rates are obtained, which are upper bounds on the minimal entanglement for the QSR task.
We denote the achievable entanglement rate for the $i$th starter by
\begin{equation} \label{eq:OldRateOfQSR}
u_i^\mathrm{QSR} = S(A_i|A_{i+ 1})_\xi + S(A_{i+ 1}|A_{i+ 2})_\xi + S(A_{i+ 2})_\xi,
\end{equation}
where additions of indices are defined modulo 3, with an offset of 1 throughout.
The relation between these rates may vary depending on the initial state $\xi$.
Thus, we take their minimum:
\begin{equation} \label{eq:QSR_old}
u_\mathrm{old}^\mathrm{QSR} \coloneqq \min\left\{u_1^\mathrm{QSR}, u_2^\mathrm{QSR}, u_3^\mathrm{QSR}\right\}.
\end{equation}

When the three parties are aware of a three-party version of the common subspace, they can reduce entanglement consumption.
To verify this, we consider a specific class of states:
\begin{eqnarray} \label{eq:Xi}
\ket{\xi}_{A_1A_2A_3R}
&=&\frac{c_0}{\sqrt{2}}\left(\ket{0010}_{A_1A_2A_3R} + \ket{0101}_{A_1A_2A_3R} \right)
+ \frac{c_1}{\sqrt{2}}\left(\ket{1202}_{A_1A_2A_3R} + \ket{2213}_{A_1A_2A_3R} \right) \\
&&
+ \frac{c_2}{\sqrt{2}}\left(\ket{3334}_{A_1A_2A_3R} + \ket{4444}_{A_1A_2A_3R} \right)
+ c_3\ket{5555}_{A_1A_2A_3R},
\end{eqnarray}
where the real numbers $c_i$ satisfy $\sum_{i=0}^3 c_i^2=1$.
The last two terms of the initial state $\xi$ do not need to be state-rotated.
The three parties can construct a subspace rotation strategy that state-rotates the first two terms of $\xi$.
Specifically, by setting a three-party common subspace as in Eq.~(\ref{eq:CommonSubspaceExample}), they can find a three-party version of the corresponding stretched state as follows:
\begin{equation}
\ket{\xi^{\text{3-str}}}_{A_1A_2A_3RA'_1A'_2A'_3}
=\ket{\xi^\text{3-com}}_{A_1A_2A_3R} \otimes\ket{3}_{A'_1} \otimes\ket{3}_{A'_2} \otimes\ket{3}_{A'_3}
+ \ket{0}_{A_1} \otimes\ket{0}_{A_2} \otimes\ket{0}_{A_3} \otimes\ket{\zeta^\text{3-unc}}_{RA'_1A'_2A'_3},
\end{equation}
where $A'_i$ is the additional quantum system of the $i$th party.
Then, the three-party versions of the common and uncommon states are given by
\begin{eqnarray}
\ket{\xi^\text{3-com}}_{A_1A_2A_3R}
&=&
\frac{c_2}{\sqrt{2}}\left(\ket{3334}_{A_1A_2A_3R} + \ket{4444}_{A_1A_2A_3R} \right)
+ c_3\ket{5555}_{A_1A_2A_3R}, \\
\ket{\xi^\text{3-unc}}_{A_1A_2A_3R}
&=&
\frac{c_0}{\sqrt{2}}\left(\ket{0010}_{A_1A_2A_3R} + \ket{0101}_{A_1A_2A_3R} \right)
+ \frac{c_1}{\sqrt{2}}\left(\ket{1202}_{A_1A_2A_3R} + \ket{2213}_{A_1A_2A_3R} \right).
\end{eqnarray}
To state-rotate the parts $A'_i$ of the stretched state $\xi^{\text{3-str}}$, they utilize a three-party version of the merge-and-merge strategy~\cite{Lee2019a}, where the parties may have and use respective QSI.
This three-party strategy proceeds as follows:
\begin{equation} \label{eq:SubspaceRotationStrategy}
1^{\mathrm{st}}\text{ party}
\xrightarrow[\quad\text{with QSI }A_2A'_2\quad]{\quad\text{Merging }A'_1\quad} 2^{\mathrm{nd}}\text{ party}
\xrightarrow[\quad\text{with QSI }A_3A'_3\quad]{\quad\text{Merging }A'_2\quad} 3^{\mathrm{rd}}\text{ party}
\xrightarrow[\quad\text{with QSI }A_1\quad]{\quad\text{Merging }A'_3\quad} 1^{\mathrm{st}}\text{ party},
\end{equation}
To obtain the final state $\xi_\mathrm{f}$ of the QSR task, the parties apply a unitary operation of the same type as the one used in Eq.~(\ref{eq:DividedFinal}).
The achievable entanglement rate for the $i$th starter is given by
\begin{equation} \label{eq:NewRateOfQSR}
v_i^\mathrm{QSR}
= S(A'_i|A_{i+ 1}A'_{i+ 1})_{\xi^{\text{3-str}}} + S(A'_{i+ 1}|A_{i+ 2}A'_{i+ 2})_{\xi^{\text{3-str}}} + S(A'_{i+ 2}|A_{i+ 3})_{\xi^{\text{3-str}}},
\end{equation}
and we define their minimum as
\begin{equation} \label{eq:QSR_new}
v_\mathrm{new}^\mathrm{QSR} \coloneqq \min\left\{v_1^\mathrm{QSR}, v_2^\mathrm{QSR}, v_3^\mathrm{QSR}\right\}.
\end{equation}

\begin{figure}[t]
\includegraphics[clip,width=.75\columnwidth]{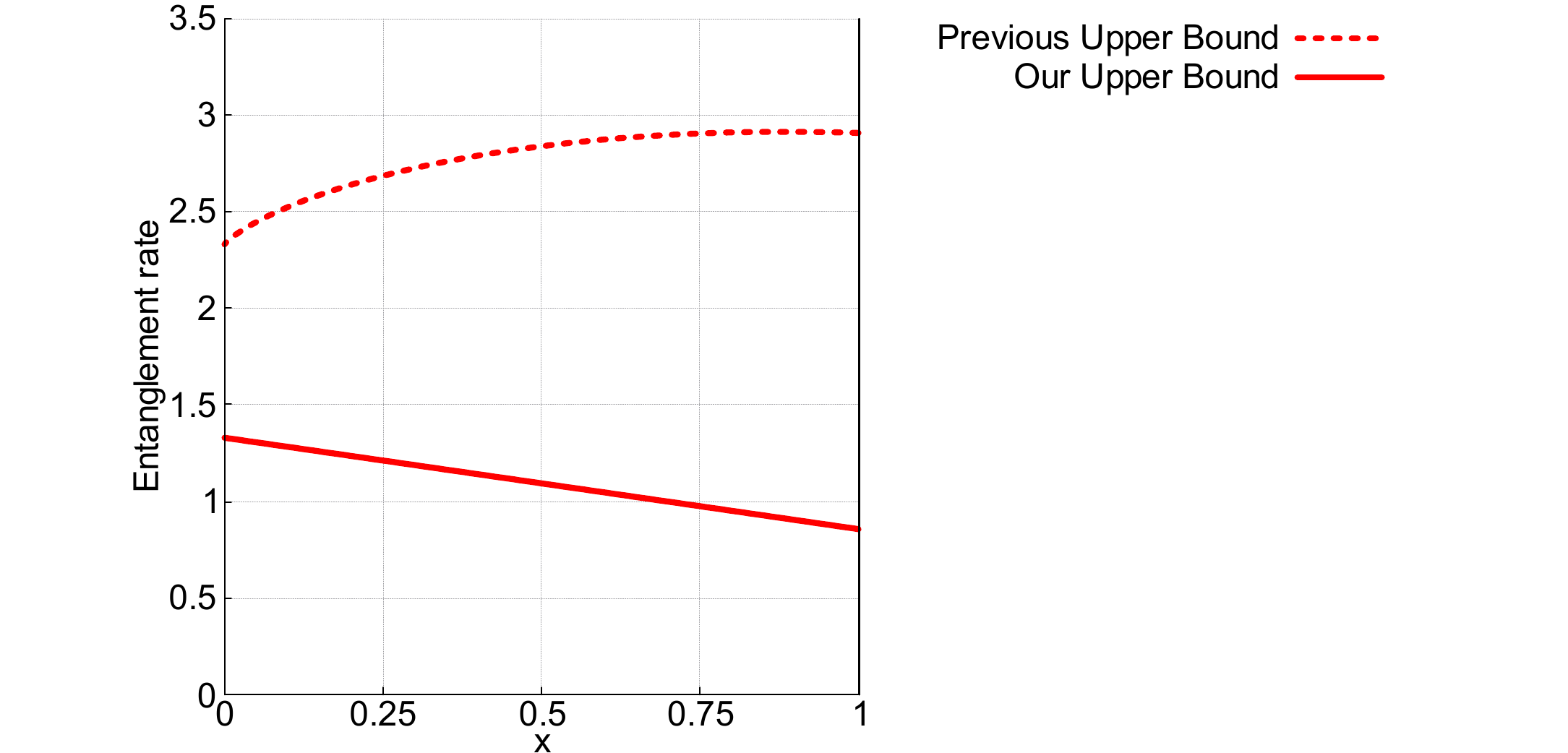}
\caption{
Comparison between our achievable bound for the QSR task and the previous one:
We parameterize the initial state $\xi$ in Eq.~(\ref{eq:Xi}) through Eq.~(\ref{eq:Parameter}) to draw graphs of the achievable bounds $v_\mathrm{new}^\mathrm{QSR}$ and $u_\mathrm{old}^\mathrm{QSR}$ in Eqs.~(\ref{eq:QSR_old}) and~(\ref{eq:QSR_new}).
With the parameter $x \in [0,1]$, we express the graph of our rate as a straight line and the previous rate as a dotted line.
The graphs verify that the three parties of the QSR task can state-rotate the initial state using less entanglement.
}
\label{fig:QSRgraphs}
\end{figure}

To compare the achievable entanglement rates $v_\mathrm{new}^\mathrm{QSR}$ and $u_\mathrm{old}^\mathrm{QSR}$, we parameterize the coefficient $c_i$ of the initial state $\xi$ as in Eq.~(\ref{eq:Parameter}).
The graphs in Fig.~\ref{fig:QSRgraphs} show that the three parties can reduce the amount of entanglement consumed in the QSR task through the subspace rotation strategy.
By generalizing the RAE task, it is theoretically possible to find an improved converse bound for the QSR task.
However, to verify this through examples, one would need to know at least one reversible transformation involving the four parties for the given initial state.
As far as we know, such a transformation is not currently known, and thus, an improved converse bound does not appear in Fig.~\ref{fig:QSRgraphs}.
In Appendix~\ref{app:Three}, we present formal definitions of a QSR protocol, its (minimal) achievable entanglement rate, and calculations for the achievable entanglement rates $v_\mathrm{new}^\mathrm{QSR}$ and $u_\mathrm{old}^\mathrm{QSR}$.

\section{Conclusion} \label{sec:Conclusion}
Although the QUI has a clear operational meaning as the least amount of entanglement required for a QSE task~\cite{Oppenheim}, its analytic closed-form expression does not derive from classical counterparts and remains an open question to date.
Note that in classical information theory, common information is represented by the mutual information $I(A;B)$ and uncommon information is its complement $H(X|Y) + H(Y|X)$.
The distinction may have its origin in the fact that quantum conditional entropies $S(A|B)$ and $S(B|A)$ can be negative while the operational definition of the QUI tells, as it is defined by a QSE task, that it cannot be negative since LOCC does not create entanglement.
The distinction of uncommon information in quantum and classical information theories may signify advantages of quantum resources.

In this work, we have presented improved lower and upper bounds for the QUI by developing two approaches: the SSE strategy and the RAE task.
These approaches are scalable and applicable to various quantum communication scenarios.
Specifically, we have applied the common subspace concept to the exact single-shot scenario and have demonstrated that non-local SWAP can be implemented with less entanglement.
Furthermore, we have considered the QSR task, where three parties transmit their respective quantum information to the next party.
By generalizing the notion of common subspace, we have confirmed the existence of a more efficient QSR protocol, enabling the three parties to communicate quantum information using less entanglement.
We anticipate that our results are applicable to various fields of quantum information, such as quantum networks~\cite{Cirac1997, Azuma2016} and distributed quantum computing~\cite{Cirac1999, Bruss2004}.

Our bounds include the previous bounds as special cases.
In the SSE strategy, if Alice and Bob are unaware of any common subspace $C$, the SSE strategy reduces to the previous merge-and-send strategy~\cite{Oppenheim}.
Additionally, in the RAE task, if any non-trivial reversible transformation expressed by Eq.~(\ref{eq:Reversible}) cannot be found, our converse bound $l_\mathrm{new}$ becomes equivalent to the previous bounds $l_1$ or $l_2$.
In this work, while we have demonstrated that our bounds are theoretically tighter than the previous bounds, the existence of the common subspace and reversible transformation has not been discussed.
From a practical perspective, it is crucial to find a common subspace and a reversible transformation for a given initial state $\psi$ to demonstrate the utility of our bounds.

In this regard, our results raise several questions that require further study.
For the SSE strategy, there has been no general method to determine whether a given initial state has a common subspace.
Furthermore, even if a common subspace exists, no algorithm to find it has been elaborated.
In addition, when two common subspaces $C_1$ and $C_2$ exist, it is unclear whether their union $C_1 \cup C_2$ forms a new common subspace of the initial state.
A related question is whether it is possible to construct the maximal common subspace that contains all common subspaces.
For the RAE task, the specific reversible transformation of Eq.~(\ref{eq:AsymptoticDecomposition}) we used as an example is not sufficient to apply to any initial states, since there exists a counterexample~\cite{Acin2003} that cannot be reversibly converted.
So, we need to to develop practical and general reversible transformations to find and use tighter converse bounds on the QUI.
We believe that answering these questions is ultimately the most effective way to find the closed-form expression of the QUI.

\section*{ACKNOWLEDGMENTS}
This research was supported by Basic Science Research Program through the National Research Foundation of Korea (NRF) funded by the Ministry of Education (No. NRF-2020R1I1A1A01058364 and No. RS-2023-00243988).
J. B. is supported by the National Research Foundation of Korea (Grant No. NRF-2021R1A2C2006309, NRF-2022M1A3C2069728), the Institute for Information \& Communication Technology Promotion (IITP) (RS-2023-00229524).
H.Y.\ acknowledges JSPS Overseas Research Fellowships, JST PRESTO Grant Number JPMJPR201A, JPMJPR23FC, JSPS KAKENHI Grant Number JP23K19970, and MEXT Quantum Leap Flagship Program (MEXT QLEAP) JPMXS0118069605, JPMXS0120351339\@.
S.L. acknowledges support from the NRF grants funded by the Ministry of Science and ICT (MSIT) (No. NRF-2022M3K2A1083859 and No. NRF-2022R1F1A1068197), and Creation of the Quantum Information Science R\&D Ecosystem (No. 2022M3H3A106307411) through the NRF funded by the MSIT.

\bibliography{UI}

\appendix

\section{Details for SSE strategy} \label{app:SES}
The details of the SSE strategy are elucidated below.
Let $\psi$ be an initial state of the QSE task representing quantum systems $A$, $B$, and $R$.
Generally, it is not necessary for the dimensions of the Hilbert spaces of systems $A$ and $B$ to be identical.
However, if their dimensions differ, for instance, if $d_A < d_B$, we can consider a Hilbert space of dimension $d_B$ that includes $\mathcal{H}_A$ as a subspace to represent system $A$.
Consequently, without loss of generality, we assume that the dimension of the Hilbert space $\mathcal{H}_A$ is identical to that of the Hilbert space $\mathcal{H}_B$, i.e., $d_A=d=d_B$.
The SSE strategy requires a common subspace $C$ with respect to the initial state $\psi$.
For convenience, we assume that the common subspace $C$ is spanned by a subset of the computational basis for $\mathcal{H}_A$:
\begin{equation}
C=\mathrm{span}\{\ket{0},\ket{1},\ldots,\ket{d_C- 1}\}.
\end{equation}

The SSE strategy consists of the following procedure.

(i) Alice and Bob apply the common unitaries defined in Definition~\ref{def:Common}, which are determined based on the common subspace $C$, to the initial state $\psi$.
The initial state then becomes the sum of a common state $\psi^\mathrm{com}$ and an uncommon state $\psi^\mathrm{unc}$:
\begin{equation} \label{eq:First}
V_A \otimes W_B \otimes \mathds{1}_{R} \ket{\psi}_{ABR} =\ket{\psi^\mathrm{com}}_{ABR} + \ket{\psi^\mathrm{unc}}_{ABR}.
\end{equation}
The common and uncommon states are represented as
\begin{eqnarray}
\ket{\psi^\mathrm{com}}_{ABR}
&=& \sum_{i,j=0}^{d_C- 1}\sum_{k=0}^{d_R- 1} c_{ijk} \ket{i}_A \otimes \ket{j}_B \otimes \ket{k}_R, \\
\ket{\psi^\mathrm{unc}}_{ABR}
&=& \sum_{i,j=d_C}^{d_A- 1}\sum_{k=0}^{d_R- 1} c_{ijk} \ket{i}_A \otimes \ket{j}_B \otimes \ket{k}_R,
\end{eqnarray}
where the coefficients $c_{ijk}$ are given by
\begin{equation}
c_{ijk}
= \bra{i}_A \otimes \bra{j}_B \otimes \bra{k}_R \left(V_A \otimes W_B \otimes \mathds{1}_{R} \right) \ket{\psi}_{ABR}.
\end{equation}
The condition in Eq.~(\ref{eq:Condition1}) becomes that
\begin{equation} \label{eq:Condition1- 1}
c_{ijk}=c_{jik}
\end{equation}
holds for each $i,j \in \{0,1, \ldots, d_C- 1\}$ and $k \in \{0,1,\ldots,d_R- 1 \}$.
The common state $\psi^\mathrm{com}$ represents a structure that does not require any entanglement for state exchange:
\begin{equation} \label{eq:NoExchange}
\ket{\psi^\mathrm{com}}_{ABR}
=\sum_{i,j=0}^{d_C- 1}\sum_{k=0}^{d_R- 1} c_{jik}\ket{i}_A \otimes\ket{j}_B \otimes \ket{k}_R
=\ket{\psi^\mathrm{com}_\mathrm{f}}_{ABR},
\end{equation}
where the first equality comes from the condition in Eq.~(\ref{eq:Condition1- 1}), and the second is obtained by exchanging the indices $i$ and $j$. Therefore, Alice and Bob do not need to state-exchange the common state.

(ii) Alice and Bob consider additional quantum systems $A'$ and $B'$.
The Hilbert spaces $\mathcal{H}_{A'}$ and $\mathcal{H}_{B'}$ are identical to $\mathcal{H}_{A}$ and $\mathcal{H}_{B}$, respectively.
To separate the common and uncommon states, they prepare pure quantum states on the systems $A'$ and $B'$, and apply the unitary $U$ on quantum systems $XX'$ defined as
\begin{equation} \label{eq:UnitaryU}
U_{XX'}
\coloneqq
\sum_{i=0}^{d_C- 1}
\ket{i}\bra{i}_X \otimes \ket{0}\bra{0}_{X'}
+ \sum_{i=d_C}^{d_{X}- 1} \left(
\ket{d_C}\bra{i}_X \otimes \ket{i}\bra{0}_{X'} + \ket{i}\bra{d_C}_X \otimes \ket{0}\bra{i}_{X'}
\right)
+ \sum_{i\neq d_C}\sum_{j\neq0}\ket{i}\bra{j}_X \otimes \ket{i}\bra{j}_{X'},
\end{equation}
where the notation $X$ can be replaced by $A$ or $B$.
This entire process is expressed by the following equation:
\begin{eqnarray}
&&\left(\ket{\psi^\mathrm{com}}_{ABR} + \ket{\psi^\mathrm{unc}}_{ABR} \right) \otimes\ket{0}_{A'} \otimes\ket{0}_{B'} \\
&&\xrightarrow{U_{AA'} \otimes U_{BB'}}
\ket{\psi^\mathrm{str}}
= \sum_{i,j=0}^{d_C- 1}\sum_{k=0}^{d_R- 1} c_{ijk} \ket{i}_A \otimes \ket{j}_B \otimes \ket{k}_R \otimes\ket{0}_{A'} \otimes\ket{0}_{B'}
+ \sum_{i,j=d_C}^{d_A- 1}\sum_{k=0}^{d_R- 1} c_{ijk} \ket{d_C}_{A} \otimes\ket{d_C}_{B} \otimes\ket{k}_R \otimes \ket{i}_{A'} \otimes \ket{j}_{B'},
\end{eqnarray}
where the stretchered state $\psi^\mathrm{str}$ is defined in Eq.~(\ref{eq:Stretched}).

(iii) Alice and Bob state-exchange their parts $A'$ and $B'$ of the stretched state $\psi^\mathrm{str}$.
In the single-shot regime, they use the quantum teleportation protocol~\cite{Bennett1993} to exchange the parts without error.
They then obtain that
\begin{equation}
\ket{\psi^\mathrm{str}_\mathrm{f}}
= \sum_{i,j=0}^{d_C- 1}\sum_{k=0}^{d_R- 1} c_{ijk} \ket{i}_A \otimes \ket{j}_B \otimes \ket{k}_R \otimes\ket{0}_{A'} \otimes\ket{0}_{B'}
+ \sum_{i,j=d_C}^{d_A- 1}\sum_{k=0}^{d_R- 1} c_{ijk} \ket{d_C}_{A} \otimes\ket{d_C}_{B} \otimes\ket{k}_R \otimes \ket{j}_{A'} \otimes \ket{i}_{B'},
\end{equation}
where the exchanged state $\psi^\mathrm{str}_\mathrm{f}$ is defined in Eq.~(\ref{eq:StretchedExchanged}).
In the asymptotic regime, they need to utilize the merge-and-merge strategy~\cite{Lee2019a} to reduce shared entanglement.
In the first merging, Alice transmits her information of the system $A'$ to Bob using the parts $BB'$ as QSI.
Then, in the second merging, Bob transmits his information of the system $B'$ to Alice using the part $A$ as QSI.
The optimal entanglement rates~\cite{Horodecki2007} for the first and second mergings are $S(A'|BB')_{\psi^\mathrm{str}}$ and $S(B'|A)_{\psi^\mathrm{str}}$, respectively.

(iv) Alice and Bob apply the inverse of the unitary $U_{XX'}$ from Eq.~(\ref{eq:UnitaryU}) to ensure the common and uncommon states represent the same systems $ABR$:
\begin{equation} \label{eq:Step4}
\ket{\psi^\mathrm{str}_\mathrm{f}} 
\xrightarrow{U^\dagger_{AA'} \otimes U^\dagger_{BB'}} \left(\ket{\psi^\mathrm{com}}_{ABR} + \ket{\psi^\mathrm{unc}_\mathrm{f}}_{ABR} \right) \otimes\ket{0}_{A'} \otimes\ket{0}_{B'}.
\end{equation}

(v) Lastly, Alice and Bob apply the inverses of the unitaries $V$ and $W$ in Eq.~(\ref{eq:First}):
\begin{equation}
\left(W_A^\dagger \otimes V_B^\dagger \otimes \mathds{1}_{R} \right)
\left( \ket{\psi^\mathrm{com}}_{ABR} + \ket{\psi^\mathrm{unc}_\mathrm{f}}_{ABR} \right)
=\ket{\psi_\mathrm{f}}_{ABR},
\end{equation}
where the equality comes from Eqs.~(\ref{eq:First}) and~(\ref{eq:NoExchange}). 

This description of the SSE strategy shows the existence of an SSE protocol.

\section{Achievability of the upper bound $u_\mathrm{new}$ on QUI} \label{app:Achievability}
To prove Theorem~\ref{thm:Achievable}, we demonstrate that the conditional entropy $S(R|A)_{\psi^\mathrm{str}}$ is an achievable entanglement rate.
It suffices to construct a sequence of QSE protocols $\mathcal{E}_n$ in Eq.~(\ref{eq:EnDomainRange}) such that
\begin{eqnarray}
\lim_{n\to\infty} \frac{1}{n}\left( \log \mathrm{Sr}[\Psi_n] - \log \mathrm{Sr}[\Phi_n] \right)
&=& S(R|A)_{\psi^\mathrm{str}}, \label{eq:C1} \\
\lim_{n\to\infty}\epsilon_n
&=&0. \label{eq:C2}
\end{eqnarray}
In this section, the symbol $\rho$ denotes input states of quantum channels.

Following the steps of the SSE strategy, we construct SSE protocols, denoted by $\mathcal{E}_n^\mathrm{SSE}$.

(i) In the first step, Alice and Bob apply the common unitaries of Definition~\ref{def:Common}.
Since unitaries are a special case of quantum channels, for each $n$, we can find a sub-protocol
\begin{equation}
\mathcal{S}_n^{(1)} \colon \mathcal{L}\left( A^{\otimes n}B^{\otimes n} \otimes D_nE_n \right) \longrightarrow \mathcal{L}\left( A^{\otimes n}B^{\otimes n} \otimes D_nE_n \right),
\end{equation}
defined as
\begin{equation}
\mathcal{S}_n^{(1)} [ \rho ]
\coloneqq
\left( V_A^{\otimes n}\otimes W_B^{\otimes n} \otimes \mathds{1}_{D_nE_n} \right)
\rho
\left( V_A^{\dagger\otimes n} \otimes W_B^{\dagger\otimes n} \otimes \mathds{1}_{D_nE_n} \right).
\end{equation}
This maps the initial state to the sum of the common and uncommon states as follows:
\begin{equation}
\left( \mathcal{S}_n^{(1)} \otimes \mathrm{id}_{R^{\otimes n}} \right) [ \psi^{\otimes n} \otimes \Psi_n^{(3)} ]
=\left(\psi^\mathrm{com}+ \psi^\mathrm{unc}\right)^{\otimes n} \otimes \Psi_n^{(3)}.
\end{equation}
The Schmidt rank of a maximally entangled state $\Psi_n^{(3)}$ is determined in the third step.

(ii) To obtain the stretched state $\psi^\mathrm{str}$ in Eq.~(\ref{eq:Stretched}), Alice and Bob employ the unitary $U_{XX'}$ in Eq.~(\ref{eq:UnitaryU}).
For each $n$, we can find a sub-protocol
\begin{equation}
\mathcal{S}_n^{(2)}
\colon
\mathcal{L}\left( A^{\otimes n}A'^{\otimes n}B^{\otimes n}B'^{\otimes n} \otimes D_nE_n \right)
\longrightarrow
\mathcal{L}\left( A^{\otimes n}A'^{\otimes n}B^{\otimes n}B'^{\otimes n} \otimes D_nE_n \right),
\end{equation}
defined as
\begin{equation}
\mathcal{S}_n^{(2)} [ \rho ]
\coloneqq
\left( U_{AA'}^{\otimes n} \otimes U_{BB'}^{\otimes n} \otimes \mathds{1}_{D_nE_n} \right)
\rho
\left( U_{AA'}^{\dagger\otimes n} \otimes U_{BB'}^{\dagger\otimes n} \otimes \mathds{1}_{D_nE_n} \right).
\end{equation}
By using the protocol $\mathcal{S}_n^{(2)}$, we obtain
\begin{equation}
\left( \mathcal{S}_n^{(2)} \otimes \mathrm{id}_{R^{\otimes n}} \right)
\left[ (\psi^\mathrm{com} + \psi^\mathrm{unc})^{\otimes n} \otimes \zeta^{\otimes n} \otimes \zeta^{\otimes n} \otimes \Psi_n^{(3)} \right]
=(\psi^\mathrm{str})^{\otimes n} \otimes \Psi_n^{(3)},
\end{equation}
where $\zeta^{\otimes n}$ denotes $n$ copies of any pure state $\zeta$ representing the quantum system $A'^{\otimes n}$ ($B'^{\otimes n}$).

(iii) Alice and Bob exchange the parts $A'$ and $B'$ of the stretchered state $\psi^\mathrm{str}$.
To this end, they employ the merge-and-merge strategy~\cite{Lee2019a}.
That is, for each $n$, they perform a protocol
\begin{equation}
\mathcal{S}_n^{(3)}
\colon
\mathcal{L}\left( A^{\otimes n}A'^{\otimes n}B^{\otimes n}B'^{\otimes n} \otimes D_nE_n \right)
\longrightarrow
\mathcal{L}\left( A^{\otimes n}A'^{\otimes n}B^{\otimes n}B'^{\otimes n} \otimes F_nG_n \right),
\end{equation}
such that 
\begin{eqnarray}
\left\|
\left( \mathcal{S}_n^{(3)} \otimes \mathrm{id}_{R^{\otimes n}} \right)
\left[ (\psi^\mathrm{str})^{\otimes n} \otimes \Psi_n^{(3)} \right]
- 
(\psi^\mathrm{str}_\mathrm{f})^{\otimes n} \otimes \Phi_n^{(3)} 
\right\|_{1}
&\le&
\epsilon_n^{(3)}, \label{eq:SubspaceCondition1} \\
\lim_{n\to\infty} \frac{1}{n}\left( \log \mathrm{Sr}\left[\Psi_n^{(3)}\right] - \log \mathrm{Sr}\left[\Phi_n^{(3)}\right] \right)
&=& S(R|A)_{\psi^\mathrm{str}}, \label{eq:SubspaceCondition2} \\
\lim_{n\to\infty}\epsilon_n^{(3)}
&=&0, \label{eq:SubspaceCondition3}
\end{eqnarray}
where $\psi^\mathrm{str}_\mathrm{f}$ is defined in Eq.~(\ref{eq:StretchedExchanged}).

(iv) For each $n$, we construct a sub-protocol
\begin{equation}
\mathcal{S}_n^{(4)}
\colon
\mathcal{L}\left( A^{\otimes n}A'^{\otimes n}B^{\otimes n}B'^{\otimes n} \otimes F_nG_n \right)
\longrightarrow
\mathcal{L}\left( A^{\otimes n}A'^{\otimes n}B^{\otimes n}B'^{\otimes n} \otimes F_nG_n \right),
\end{equation}
representing the unitaries $U_{XX'}^\dagger$ in Eq.~(\ref{eq:DividedFinal}), defined as
\begin{equation}
\mathcal{S}_n^{(4)} [ \rho ]
\coloneqq
\left( U_{AA'}^{\dagger\otimes n} \otimes U_{BB'}^{\dagger\otimes n} \otimes \mathds{1}_{F_nG_n} \right)
\rho
\left( U_{AA'}^{\otimes n} \otimes U_{BB'}^{\otimes n} \otimes \mathds{1}_{F_nG_n} \right).
\end{equation}
We then obtain 
\begin{equation}
\left( \mathcal{S}_n^{(4)} \otimes \mathrm{id}_{R^{\otimes n}} \right)
\left[ (\psi^\mathrm{str}_\mathrm{f})^{\otimes n} \otimes \Phi_n^{(3)} \right]
=\left( \psi^\mathrm{com} + \psi^\mathrm{unc}_\mathrm{f} \right)^{\otimes n} \otimes \zeta^{\otimes n} \otimes \zeta^{\otimes n} \otimes \Phi_n^{(3)}.
\end{equation}

(v) In the last step, for each $n$, we construct a sub-protocol
\begin{equation}
\mathcal{S}_n^{(5)}
\colon
\mathcal{L}\left( A^{\otimes n} B^{\otimes n} \otimes F_nG_n \right)
\longrightarrow
\mathcal{L}\left( A^{\otimes n} B^{\otimes n} \otimes F_nG_n \right),
\end{equation}
defined as
\begin{equation}
\mathcal{S}_n^{(5)} [ \rho ]
\coloneqq
\left( V_A^{\dagger\otimes n} \otimes W_B^{\dagger\otimes n} \otimes \mathds{1}_{F_nG_n} \right)
\rho
\left( V_A^{\otimes n} \otimes W_B^{\otimes n} \otimes \mathds{1}_{F_nG_n} \right).
\end{equation}
This sub-protocol completes the SSE strategy:
\begin{equation}
\left( \mathcal{S}_n^{(5)} \otimes \mathrm{id}_{R^{\otimes n}} \right)
\left[ \left(\psi^\mathrm{com} + \psi^\mathrm{unc}_\mathrm{f} \right)^{\otimes n} \otimes \Phi_n^{(3)} \right]
=\psi_\mathrm{f}^{\otimes n} \otimes \Phi_n^{(3)}.
\end{equation}

Note that all previously mentioned sub-protocols are LOCC.
In particular, only the third-stage protocol $\mathcal{S}_n^{(3)}$ handles the entanglement shared between Alice and Bob.
By composing these sub-protocols, we construct a SSE protocol $\mathcal{E}_n^\mathrm{SSE}$ as 
\begin{equation}
\mathcal{E}_n^\mathrm{SSE}[\rho]
\coloneqq
\mathcal{S}_n^{(5)}\left[
\mathrm{Tr}_{A'^{\otimes n}B'^{\otimes n}}\left[
\mathcal{S}_n^{(4)}\left[
\mathcal{S}_n^{(3)}\left[
\mathcal{S}_n^{(2)}\left[
\mathcal{S}_n^{(1)}\left[ \rho \right]\otimes \zeta^{\otimes n} \otimes \zeta^{\otimes n}
\right] \right] \right] \right] \right].
\end{equation}
The SSE protocol has the same domain and range as the QSE protocol in Eq.~(\ref{eq:EnDomainRange}).
It maps an input $\psi^{\otimes n} \otimes \Psi_n^{(3)}$ to an output $\psi_\mathrm{f}^{\otimes n} \otimes \Phi_n^{(3)}$ with an error $\epsilon_n^{(3)}$:
\begin{eqnarray}
&&\left\|
\left( \mathcal{E}_n^\mathrm{SSE} \otimes \mathrm{id}_{\bar{R}} \right)
\left[ \psi^{\otimes n} \otimes \Psi_n^{(3)} \right]
- 
\psi_\mathrm{f}^{\otimes n} \otimes \Phi_n^{(3)}
\right\|_{1} \\
&&=
\left\|
\mathcal{S}_n^{(5)}\otimes \mathrm{id}_{\bar{R}}\left[
\mathrm{Tr}_{A'^{\otimes n}B'^{\otimes n}}\left[
\mathcal{S}_n^{(4)}\otimes \mathrm{id}_{\bar{R}}\left[
\mathcal{S}_n^{(3)}\otimes \mathrm{id}_{\bar{R}}\left[
\mathcal{S}_n^{(2)}\otimes \mathrm{id}_{\bar{R}}\left[
\mathcal{S}_n^{(1)}\otimes \mathrm{id}_{\bar{R}}\left[ \psi^{\otimes n} \otimes \Psi_n^{(3)} \right]\otimes \zeta^{\otimes n} \otimes \zeta^{\otimes n}
\right] \right] \right] \right] \right]
- 
\psi_\mathrm{f}^{\otimes n} \otimes \Phi_n^{(3)}
\right\|_{1} \\
&&=
\left\|
\mathcal{S}_n^{(5)}\otimes \mathrm{id}_{\bar{R}}\left[
\mathrm{Tr}_{A'^{\otimes n}B'^{\otimes n}}\left[
\mathcal{S}_n^{(4)}\otimes \mathrm{id}_{\bar{R}}\left[
\mathcal{S}_n^{(3)}\otimes \mathrm{id}_{\bar{R}}\left[
\mathcal{S}_n^{(2)}\otimes \mathrm{id}_{\bar{R}}\left[
\left(\psi^\mathrm{com} + \psi^\mathrm{unc} \right)^{\otimes n} \otimes \zeta^{\otimes n} \otimes \zeta^{\otimes n} \otimes \Psi_n^{(3)}
\right] \right] \right] \right] \right]
- 
\psi_\mathrm{f}^{\otimes n} \otimes \Phi_n^{(3)}
\right\|_{1} \\
&&=
\left\|
\mathcal{S}_n^{(5)}\otimes \mathrm{id}_{\bar{R}}\left[
\mathrm{Tr}_{A'^{\otimes n}B'^{\otimes n}}\left[
\mathcal{S}_n^{(4)}\otimes \mathrm{id}_{\bar{R}}\left[
\mathcal{S}_n^{(3)}\otimes \mathrm{id}_{\bar{R}}\left[
\left( \psi^\mathrm{str} \right)^{\otimes n} \otimes \Psi_n^{(3)}
\right] \right] \right] \right]
- 
\psi_\mathrm{f}^{\otimes n} \otimes \Phi_n^{(3)}
\right\|_{1} \\
&&=
\left\|
\mathcal{S}_n^{(5)}\otimes \mathrm{id}_{\bar{R}}\left[
\mathrm{Tr}_{A'^{\otimes n}B'^{\otimes n}}\left[
\mathcal{S}_n^{(4)}\otimes \mathrm{id}_{\bar{R}}\left[
\mathcal{S}_n^{(3)}\otimes \mathrm{id}_{\bar{R}}\left[
\left( \psi^\mathrm{str} \right)^{\otimes n} \otimes \Psi_n^{(3)}
\right] \right] \right] \right]
- 
\mathcal{S}_n^{(5)} \otimes \mathrm{id}_{\bar{R}} \left[ \left(\psi^\mathrm{com} + \psi^\mathrm{unc}_\mathrm{f} \right)^{\otimes n} \otimes \Phi_n^{(3)} \right]
\right\|_{1} \\
&&\le
\left\|
\mathrm{Tr}_{A'^{\otimes n}B'^{\otimes n}}\left[
\mathcal{S}_n^{(4)}\otimes \mathrm{id}_{\bar{R}}\left[
\mathcal{S}_n^{(3)}\otimes \mathrm{id}_{\bar{R}}\left[
\left( \psi^\mathrm{str} \right)^{\otimes n} \otimes \Psi_n^{(3)}
\right] \right] \right]
- 
\left( \psi^\mathrm{com} + \psi^\mathrm{unc}_\mathrm{f} \right)^{\otimes n} \otimes \Phi_n^{(3)}
\right\|_{1} \\
&&\le
\left\|
\mathcal{S}_n^{(4)}\otimes \mathrm{id}_{\bar{R}}\left[
\mathcal{S}_n^{(3)}\otimes \mathrm{id}_{\bar{R}}\left[
\left( \psi^\mathrm{str} \right)^{\otimes n} \otimes \Psi_n^{(3)}
\right] \right]
- 
\left( \psi^\mathrm{com} + \psi^\mathrm{unc}_\mathrm{f} \right)^{\otimes n} \otimes \zeta^{\otimes n} \otimes \zeta^{\otimes n} \otimes \Phi_n^{(3)}
\right\|_{1} \\
&&=
\left\|
\mathcal{S}_n^{(4)}\otimes \mathrm{id}_{\bar{R}}\left[
\mathcal{S}_n^{(3)}\otimes \mathrm{id}_{\bar{R}}\left[
\left( \psi^\mathrm{str} \right)^{\otimes n} \otimes \Psi_n^{(3)}
\right] \right]
- 
\mathcal{S}_n^{(4)} \otimes \mathrm{id}_{\bar{R}} \left[ \left(\psi^\mathrm{str}_\mathrm{f} \right)^{\otimes n} \otimes \Phi_n^{(3)} \right]
\right\|_{1} \\
&&\le
\left\|
\mathcal{S}_n^{(3)}\otimes \mathrm{id}_{\bar{R}}\left[
\left( \psi^\mathrm{str} \right)^{\otimes n} \otimes \Psi_n^{(3)}
\right]
- 
\left(\psi^\mathrm{str}_\mathrm{f} \right)^{\otimes n} \otimes \Phi_n^{(3)}
\right\|_{1} \\
&&\le
\epsilon_n^{(3)},
\end{eqnarray}
where the notation $\mathrm{id}_{\bar{R}}$ denotes $\mathrm{id}_{R^{\otimes n}}$.
The first three inequalities arise from the monotonicity of the trace distance~\cite{Wilde2013}, and the last inequality follows from Eq.~(\ref{eq:SubspaceCondition1}).
Eq.~(\ref{eq:SubspaceCondition3}) implies that the error $\epsilon_n^{(3)}$ of the protocol $\mathcal{E}_n^\mathrm{SSE}$ converges to zero as $n$ tends to infinity.
Since $\Psi_n^{(3)}$ and $\Phi_n^{(3)}$ are the only entanglement involved in the SSE strategy, the conditional entropy $S(R|A)_{\psi^\mathrm{str}}$ in Eq.~(\ref{eq:C1}) becomes an achievable entanglement rate of the QSE task.
Consequently, we have constructed QSE protocols with an achievable entanglement rate of $S(R|A)_{\psi^\mathrm{str}}$.

\section{Derivation of the converse bound $l_\mathrm{new}$ on QUI} \label{app:Converse}
To prove Theorem~\ref{thm:Converse}, we first describe the RAE task and its achievable entanglement rate.
Then, we derive a converse bound on the minimal achievable entanglement rate for the RAE task and demonstrate that this converse bound serves as a converse bound on the QUI.

\subsection{Details for RAE task}
We describe an RAE protocol $\mathcal{E}_n^{\mathrm{RAE}}$ according to the steps of the RAE task.

(i) 
Alice, Bob, and the referee perform a reversible transformation $\Lambda$ as described in Eq.~(\ref{eq:Reversible}) to decompose the initial state $\psi$.
For each $n$, they perform a three-party transformation
\begin{equation}
\Lambda_n\colon
\mathcal{L}\left(
A^{\otimes n} \otimes B^{\otimes n} \otimes R^{\otimes n}
\right)
\longrightarrow
\mathcal{L}\left(
A_2^{ \otimes \lfloor r_2 n \rfloor}A_1^{ \otimes \lfloor r_1 n \rfloor}A_3^{ \otimes \lfloor r_4 n \rfloor}
\otimes
B_1^{ \otimes \lfloor r_3 n \rfloor}B_2^{ \otimes \lfloor r_2 n \rfloor}B_3^{ \otimes \lfloor r_4 n \rfloor}
\otimes
R_1^{ \otimes \lfloor r_1 n \rfloor}R_2^{ \otimes \lfloor r_3 n \rfloor}R_3^{ \otimes \lfloor r_4 n \rfloor}R_4^{ \otimes \lfloor r_4 n \rfloor}
\right)
\end{equation}
such that
\begin{eqnarray}
\left\|
\Lambda_n \left[ \psi^{\otimes n} \right]
- 
\psi_n^\mathrm{dec}
\right\|_{1}
&\le&
\epsilon_n^{(1)}, \\
\lim_{n\to\infty}\epsilon_n^{(1)}&=&0, \label{eq:Error1}
\end{eqnarray}
where the decomposed state $\psi_n^\mathrm{dec}$ is presented in Eq.~(\ref{eq:Reversible}).

(ii) For each $n$, the referee transmits the parts $R_1^{\otimes n}R_3^{\otimes n}$ and $R_2^{\otimes n}R_4^{\otimes n}$ to Alice and Bob, respectively.
The entanglement used by the referee is not counted, as the QUI quantifies the entanglement used by Alice and Bob.

(iii) Alice and Bob state-exchange their parts $A_i$ and $B_i$ of the decomposed state $\psi_n^\mathrm{dec}$.
This is represented as a quantum channel $\mathcal{\bar{E}}_n$.
Specifically, for each $n$, they apply a three-party protocol
\begin{eqnarray}
&&\mathcal{\bar{E}}_n\colon
\mathcal{L}\left(
A_2^{ \otimes \lfloor r_2 n \rfloor}A_1^{ \otimes \lfloor r_1 n \rfloor}A_3^{ \otimes \lfloor r_4 n \rfloor}
R_1^{ \otimes \lfloor r_1 n \rfloor}R_3^{ \otimes \lfloor r_4 n \rfloor}
R_2^{ \otimes \lfloor r_3 n \rfloor}R_4^{ \otimes \lfloor r_4 n \rfloor}
B_1^{ \otimes \lfloor r_3 n \rfloor}B_2^{ \otimes \lfloor r_2 n \rfloor}B_3^{ \otimes \lfloor r_4 n \rfloor}
\otimes D_nE_n
\right) \nonumber \\
&&~\quad\quad
\longrightarrow
\mathcal{L}\left(
A_1^{ \otimes \lfloor r_3 n \rfloor}A_2^{ \otimes \lfloor r_2 n \rfloor}A_3^{ \otimes \lfloor r_4 n \rfloor}
R_1^{ \otimes \lfloor r_1 n \rfloor}R_3^{ \otimes \lfloor r_4 n \rfloor}
R_2^{ \otimes \lfloor r_3 n \rfloor}R_4^{ \otimes \lfloor r_4 n \rfloor}
B_1^{ \otimes \lfloor r_1 n \rfloor}B_2^{ \otimes \lfloor r_2 n \rfloor}B_3^{ \otimes \lfloor r_4 n \rfloor}
\otimes F_nG_n
\right), \label{eq:e_bar}
\end{eqnarray}
such that
\begin{eqnarray} \label{eq:ErrorRn}
\left\|
\mathcal{\bar{E}}_n \left[ \psi_n^\mathrm{dec} \otimes \Psi'_n \right]
- 
\psi_{n,\mathrm{f}}^\mathrm{dec}
\otimes \Phi'_n
\right\|_{1}
&\le&
\epsilon_n^{(2)}, \label{eq:thirdStageProtocol} \\
\lim_{n\to\infty}\epsilon_n^{(2)}&=&0, \label{eq:Error2}
\end{eqnarray}
where $D_nF_n$ and $E_nG_n$ are quantum systems of Alice and Bob, and $\Psi'_n$ and $\Phi'_n$ indicate pure maximally entangled states on $D_nE_n$ and $F_nG_n$ with Schmidt ranks $\mathrm{Sr}[\Psi'_n]$ and $\mathrm{Sr}[\Phi'_n]$.
The output state $\psi_{n,\mathrm{f}}^\mathrm{dec}$ is presented in Eq.~(\ref{eq:DecomposedFinal}).
Note that at the beginning and end of the protocol $\mathcal{\bar{E}}_n$, Alice and Bob share all parts of the decomposed state $\psi_n^\mathrm{dec}$, while the referee has none.
However, during the protocol, Alice and Bob can send $R_i$ to the referee and receive it back.
The entanglement used in these transmissions is omitted in Eqs.~(\ref{eq:e_bar}) and~(\ref{eq:ErrorRn}).

(iv) For each $n$, Alice and Bob return the parts $R_i^{\otimes n}$ to the referee.
The entanglement used in this process is not considered.

(v) Recall that the transformation $\Lambda$ is reversible.
So, for each $n$, there exists a three-party transformation
\begin{equation}
\Lambda_{n,\mathrm{f}}^{- 1}\colon
\mathcal{L}\left(
A_1^{ \otimes \lfloor r_3 n \rfloor}A_2^{ \otimes \lfloor r_2 n \rfloor}A_3^{ \otimes \lfloor r_4 n \rfloor}
\otimes
R_1^{ \otimes \lfloor r_1 n \rfloor}R_2^{ \otimes \lfloor r_3 n \rfloor}R_3^{ \otimes \lfloor r_4 n \rfloor}R_4^{ \otimes \lfloor r_4 n \rfloor}
\otimes
B_1^{ \otimes \lfloor r_1 n \rfloor}B_2^{ \otimes \lfloor r_2 n \rfloor}B_3^{ \otimes \lfloor r_4 n \rfloor}
\right)
\longrightarrow
\mathcal{L}\left(
A^{\otimes n} \otimes R^{\otimes n} \otimes B^{\otimes n}
\right)
\end{equation}
such that
\begin{eqnarray}
\left\|
\Lambda_{n,\mathrm{f}}^{- 1} \left[ \psi_{n,\mathrm{f}}^\mathrm{dec} \right]
- 
\psi_\mathrm{f}^{\otimes n}
\right\|_{1}
&\le&
\epsilon_n^{(3)}, \\
\lim_{n\to\infty}\epsilon_n^{(3)}&=&0, \label{eq:Error3}
\end{eqnarray}
where the role of Alice (Bob) in $\Lambda_{n,\mathrm{f}}^{- 1}$ is the role of Bob (Alice) in $\Lambda_n^{- 1}$, and the role of the referee in $\Lambda_{n,\mathrm{f}}^{- 1}$ is the same as that of the referee in $\Lambda_n^{- 1}$.
This enables the three parties to transform $\psi_{n,\mathrm{f}}^\mathrm{dec}$ into the final state $\psi_\mathrm{f}^{\otimes n}$.

Recall that in the second, third, and fourth steps, we do not consider the entanglement spent by the referee and the others.
In particular, the quantum channels corresponding to the second and fourth steps simply redistribute the quantum states of the referee.
For this reason, these quantum channels are not explicitly expressed.

As a result, an RAE protocol $\mathcal{E}_n^{\mathrm{RAE}}$ is represented as
\begin{equation}
\mathcal{E}_n^{\mathrm{RAE}}[\rho]
=\left(
\left( \Lambda_{n,\mathrm{f}}^{- 1} \otimes \mathrm{id}_{F_nG_n} \right) \circ \mathcal{\bar{E}}_n \circ \left( \Lambda_n \otimes \mathrm{id}_{D_nE_n} \right)
\right)[\rho].
\end{equation}
It performs the RAE task for the initial state $\psi$, i.e.,
\begin{eqnarray}
&&\left\|
\mathcal{E}_n^{\mathrm{RAE}} \left[ \psi^{\otimes n} \otimes \Psi'_n \right]
- 
\psi_\mathrm{f}^{\otimes n} \otimes \Phi'_n
\right\|_{1} \\
&&=
\left\|
\left( \Lambda_{n,\mathrm{f}}^{- 1} \otimes \mathrm{id}_{F_nG_n} \right)
\left[\mathcal{\bar{E}}_n \left[ \Lambda_n[\psi^{\otimes n}] \otimes \Psi'_n \right] \right]
- 
\psi_\mathrm{f}^{\otimes n} \otimes \Phi'_n
\right\|_{1} \\
&&\le
\left\|
\left( \Lambda_{n,\mathrm{f}}^{- 1} \otimes \mathrm{id}_{F_nG_n} \right)
\left[\mathcal{\bar{E}}_n \left[ \Lambda_n[\psi^{\otimes n}] \otimes \Psi'_n \right] \right]
- 
\Lambda_{n,\mathrm{f}}^{- 1} \left[ \psi_{n,\mathrm{f}}^\mathrm{dec} \right] \otimes \Phi'_n
\right\|_{1}
+ 
\left\|
\Lambda_{n,\mathrm{f}}^{- 1} \left[ \psi_{n,\mathrm{f}}^\mathrm{dec} \right] \otimes \Phi'_n
- 
\psi_\mathrm{f}^{\otimes n} \otimes \Phi'_n
\right\|_{1} \\
&&\le
\left\|
\left( \Lambda_{n,\mathrm{f}}^{- 1} \otimes \mathrm{id}_{F_nG_n} \right)
\left[\mathcal{\bar{E}}_n \left[ \Lambda_n[\psi^{\otimes n}] \otimes \Psi'_n \right] \right]
- 
\Lambda_{n,\mathrm{f}}^{- 1} \left[ \psi_{n,\mathrm{f}}^\mathrm{dec} \right] \otimes \Phi'_n
\right\|_{1}
+ \epsilon_n^{(3)} \\
&&\le
\left\|
\mathcal{\bar{E}}_n \left[ \Lambda_n[\psi^{\otimes n}] \otimes \Psi'_n \right]
- 
\psi_{n,\mathrm{f}}^\mathrm{dec} \otimes \Phi'_n
\right\|_{1}
+ \epsilon_n^{(3)} \\
&&\le
\left\|
\mathcal{\bar{E}}_n \left[ \Lambda_n[\psi^{\otimes n}] \otimes \Psi'_n \right]
- \mathcal{\bar{E}}_n \left[ \psi_n^\mathrm{dec} \otimes \Psi'_n \right]
\right\|_{1}
+ 
\left\|
\mathcal{\bar{E}}_n \left[ \psi_n^\mathrm{dec} \otimes \Psi'_n \right]
- 
\psi_{n,\mathrm{f}}^\mathrm{dec} \otimes \Phi'_n
\right\|_{1}
+ \epsilon_n^{(3)} \\
&&\le
\left\|
\mathcal{\bar{E}}_n \left[ \Lambda_n[\psi^{\otimes n}] \otimes \Psi'_n \right]
- \mathcal{\bar{E}}_n \left[ \psi_n^\mathrm{dec} \otimes \Psi'_n \right]
\right\|_{1}
+ 
\epsilon_n^{(2)}
+ \epsilon_n^{(3)} \\
&&\le
\left\|
 \Lambda_n [\psi^{\otimes n}]- \psi_n^\mathrm{dec} 
\right\|_{1}
+ 
\epsilon_n^{(2)}
+ \epsilon_n^{(3)} \\
&&\le
\epsilon_n^{(1)}
+ \epsilon_n^{(2)}
+ \epsilon_n^{(3)},
\end{eqnarray}
where the inequalities come from the triangle property and the monotonicity of the trace distance~\cite{Wilde2013} with Eqs.~(\ref{eq:Error1}),~(\ref{eq:Error2}), and~(\ref{eq:Error3}). The sum of errors converges to zero as $n$ tends to infinity.

In the RAE protocols $\mathcal{E}_n^{\mathrm{RAE}}$, we only consider shared entanglement used in the three-party protocols $\mathcal{\bar{E}}_n$ to define an achievable entanglement rate for the RAE task as
\begin{equation}
r[\Lambda,\mathcal{\bar{E}}_n]
\coloneqq
\lim_{n\to\infty} \frac{\log \mathrm{Sr}\left[ \Psi'_n \right] - \log \mathrm{Sr}\left[ \Phi'_n \right]}{n}.
\end{equation}
The minimal achievable entanglement rate of the RAE task is defined as
\begin{equation}
\Upsilon^{\mathrm{RAE}}(A;B)_{\Lambda[\psi]} \coloneqq \inf\left\{r[\Lambda,\mathcal{\bar{E}}_n] : r[\Lambda,\mathcal{\bar{E}}_n]\text{ is the achievable entanglement rate of the RAE task for } \psi \right\},
\end{equation}
where the infimum is taken over all the third-stage protocols $\mathcal{\bar{E}}_n$ in Eq.~(\ref{eq:thirdStageProtocol}).

\subsection{Converse bound $l_\mathrm{new}$ for RAE task}
We show that any achievable entanglement rate $r[\Lambda,\mathcal{\bar{E}}_n]$ is lower-bounded by $l_\mathrm{new}$.
Let $e_n^{\mathrm{bef}}$ and $e_n^{\mathrm{aft}}$ be the amounts of entanglement between Alice and Bob before and after performing $\mathcal{\bar{E}}_n$.
Since the amount of entanglement between two parties cannot increase on average via LOCC~\cite{Bennett1996a}, we obtain that, for each $n$,
\begin{equation} \label{eq:LOCCmonotone}
e_n^{\mathrm{bef}} \ge e_n^{\mathrm{aft}},
\end{equation}
where the amounts are represented as
\begin{eqnarray}
e_n^{\mathrm{bef}}
&=&
S\left(A_1^{ \otimes \lfloor r_1 n \rfloor}A_2^{ \otimes \lfloor r_2 n \rfloor}A_3^{ \otimes \lfloor r_4 n \rfloor}
R_1^{ \otimes \lfloor r_1 n \rfloor}R_3^{ \otimes \lfloor r_4 n \rfloor}D_n
\right)_{\psi_n^\mathrm{dec} \otimes \Psi'_n}, \\
e_n^{\mathrm{aft}}
&=&
S\left(A_1^{ \otimes \lfloor r_3 n \rfloor}A_2^{ \otimes \lfloor r_2 n \rfloor}A_3^{ \otimes \lfloor r_4 n \rfloor}
R_1^{ \otimes \lfloor r_1 n \rfloor}R_3^{ \otimes \lfloor r_4 n \rfloor}F_n
\right)_{\mathcal{\bar{E}}_n [ \psi_n^\mathrm{dec} \otimes \Psi'_n ]}.
\end{eqnarray}
The amount of entanglement before the protocol is calculated as
\begin{equation}
e_n^{\mathrm{bef}}
=\lfloor r_2 n \rfloor S(A_2)_{\phi^\mathrm{b}} 
+ \lfloor r_4 n \rfloor S(A_3R_3)_{\phi^\mathrm{c}} 
+ \log \mathrm{Sr}[\Psi'_n].
\end{equation}
It is not easy to calculate $e_n^{\mathrm{aft}}$ directly; instead, we calculate its lower bound. We apply the monotonicity of the trace distance~\cite{Wilde2013} to the inequality in Eq.~(\ref{eq:ErrorRn}), and obtain 
\begin{equation}
\epsilon'_n\coloneqq
\frac{1}{2}
\left\|
\mathrm{Tr}_{R_2^{ \otimes \lfloor r_3 n \rfloor}R_4^{ \otimes \lfloor r_4 n \rfloor}B_1^{ \otimes \lfloor r_1 n \rfloor}B_2^{ \otimes \lfloor r_2 n \rfloor}B_3^{ \otimes \lfloor r_4 n \rfloor}G_n}
\left[
\mathcal{\bar{E}}_n [ \psi_n^\mathrm{dec} \otimes \Psi'_n ]
\right]
- 
\mathrm{Tr}_{R_2^{ \otimes \lfloor r_3 n \rfloor}R_4^{ \otimes \lfloor r_4 n \rfloor}B_1^{ \otimes \lfloor r_1 n \rfloor}B_2^{ \otimes \lfloor r_2 n \rfloor}B_3^{ \otimes \lfloor r_4 n \rfloor}G_n}
\left[
\psi_{n,\mathrm{f}}^\mathrm{dec} \otimes \Phi'_n
\right]
\right\|_{1}
\le
\epsilon_n^{(2)}.
\end{equation}
The continuity of the von Neumann entropy~\cite{Wilde2013,Fannes1973,Audenaert2007} implies
\begin{eqnarray}
&&\left|
e_n^{\mathrm{aft}}
- S\left(A_1^{ \otimes \lfloor r_3 n \rfloor}A_2^{ \otimes \lfloor r_2 n \rfloor}A_3^{ \otimes \lfloor r_4 n \rfloor}
R_1^{ \otimes \lfloor r_1 n \rfloor}R_3^{ \otimes \lfloor r_4 n \rfloor}F_n\right)_{\psi_{n,\mathrm{f}}^\mathrm{dec} \otimes \Phi'_n}
\right| \\
&&\le \epsilon'_n \log\left( d_{A_1^{ \otimes \lfloor r_3 n \rfloor}A_2^{ \otimes \lfloor r_2 n \rfloor}A_3^{ \otimes \lfloor r_4 n \rfloor}
R_1^{ \otimes \lfloor r_1 n \rfloor}R_3^{ \otimes \lfloor r_4 n \rfloor}F_n}- 1 \right) + h(\epsilon'_n) \\
&&\le \epsilon'_n
\left( r_3 n \log d_{A_1}
+ r_2 n \log d_{A_2}
+ r_1 n \log d_{R_1}
+ r_4 n \log (d_{A_3}d_{R_3})
+ \log \mathrm{Sr}[\Phi'_n]
\right) + h(\epsilon'_n),
\end{eqnarray}
where $h(\cdot)$ is the binary entropy~\cite{Cover2006}. The additivity of the von Neumann entropy~\cite{Wilde2013} implies
\begin{eqnarray}
&& S\left(A_1^{ \otimes \lfloor r_3 n \rfloor}A_2^{ \otimes \lfloor r_2 n \rfloor}A_3^{ \otimes \lfloor r_4 n \rfloor}
R_1^{ \otimes \lfloor r_1 n \rfloor}R_3^{ \otimes \lfloor r_4 n \rfloor}F_n\right)_{\psi_{n,\mathrm{f}}^\mathrm{dec} \otimes \Phi'_n} \\
&&=
\lfloor r_3 n \rfloor S(A_1)_{\phi^\mathrm{r}_\mathrm{f}} 
+ \lfloor r_2 n \rfloor S(A_2)_{\phi^\mathrm{b}_\mathrm{f}} 
+ \lfloor r_4 n \rfloor S(A_3R_3)_{\phi^\mathrm{c}_\mathrm{f}} 
+ \lfloor r_1 n \rfloor S(R_1)_{\phi^\mathrm{l}_\mathrm{f}} 
+ \log \mathrm{Sr}[\Phi'_n] \\
&&=
\lfloor r_3 n \rfloor S(B_1)_{\phi^\mathrm{r}} 
+ \lfloor r_2 n \rfloor S(A_2)_{\phi^\mathrm{b}} 
+ \lfloor r_4 n \rfloor S(B_3R_3)_{\phi^\mathrm{c}} 
+ \lfloor r_1 n \rfloor S(A_1)_{\phi^\mathrm{l}} 
+ \log \mathrm{Sr}[\Phi'_n],
\end{eqnarray}
where the state $\phi^\mathrm{r}_\mathrm{f}$ denotes the exchanged state of $\phi^\mathrm{r}$, and the other exchanged states are also represented using the same subscript.
Then, Eq.~(\ref{eq:LOCCmonotone}) becomes
\begin{eqnarray}
&& \lfloor r_2 n \rfloor S(A_2)_{\phi^\mathrm{b}} 
+ \lfloor r_4 n \rfloor S(A_3R_3)_{\phi^\mathrm{c}} 
+ \log \mathrm{Sr}[\Psi'_n] \\
&& \ge
\lfloor r_3 n \rfloor S(B_1)_{\phi^\mathrm{r}} 
+ \lfloor r_2 n \rfloor S(A_2)_{\phi^\mathrm{b}} 
+ \lfloor r_4 n \rfloor S(B_3R_3)_{\phi^\mathrm{c}} 
+ \lfloor r_1 n \rfloor S(A_1)_{\phi^\mathrm{l}} 
+ \log \mathrm{Sr}[\Phi'_n] \\
&& \quad
- \epsilon'_n
\left( r_3 n \log d_{A_1}
+ r_2 n \log d_{A_2}
+ r_1 n \log d_{R_1}
+ r_4 n \log (d_{A_3}d_{R_3})
+ \log \mathrm{Sr}[\Phi'_n]
\right) - h(\epsilon'_n).
\end{eqnarray}
Since the above inequality holds for each $n$, we obtain
\begin{eqnarray}
\frac{\log \mathrm{Sr}[\Psi'_n] - \log \mathrm{Sr}[\Phi'_n]}{n}
&\ge& \frac{\lfloor r_1 n \rfloor}{n} S(A_1)_{\phi^\mathrm{l}}
+ \frac{\lfloor r_3 n \rfloor}{n} S(B_1)_{\phi^\mathrm{r}}
+ \frac{\lfloor r_4 n \rfloor}{n} \left( S(B_3R_3)_{\phi^\mathrm{c}} - S(A_3R_3)_{\phi^\mathrm{c}} \right) \\
&&
- \epsilon'_n
\left( r_3 \log d_{A_1}
+ r_2 \log d_{A_2}
+ r_1 \log d_{R_1}
+ r_4 \log (d_{A_3}d_{R_3})
+ \frac{1}{n} \log \mathrm{Sr}[\Phi'_n]
\right) - \frac{h(\epsilon'_n)}{n}. \label{eq:TwoTerms}
\end{eqnarray}
The last two terms converge to zero as $n$ approaches infinity. It follows that $l_\mathrm{new}[\Lambda] \le r[\Lambda,\mathcal{\bar{E}}_n]$. This inequality holds for any protocols $\mathcal{\bar{E}}_n$. Therefore, $l_\mathrm{new}[\Lambda]$ becomes a converse bound on $\Upsilon^{\mathrm{RAE}}(A;B)_{\Lambda[\psi]}$, i.e.,
\begin{equation} \label{eq:app:RAENeed}
l_\mathrm{new}[\Lambda] \le \Upsilon^{\mathrm{RAE}}(A;B)_{\Lambda[\psi]}.
\end{equation}

\subsection{Converse bound $l_\mathrm{new}$ for QSE task}
To prove Theorem~\ref{thm:Converse}, we demonstrate that 
\begin{equation} \label{eq:app:Need}
\Upsilon^{\mathrm{RAE}}(A;B)_{\Lambda[\psi]}
\le
\Upsilon(A;B)_\psi,
\end{equation}
where $\Lambda$ denotes the reversible transformation in Eq.~(\ref{eq:Reversible}), and $\Upsilon^{\mathrm{RAE}}$ and $\Upsilon$ are the minimal achievable entanglement rates of the RAE task and the QSE task, respectively.
It suffices to show that any achievable entanglement rate of the QSE task is also achievable for the RAE task.

Let $r'$ be any achievable entanglement rate of the QSE task for the initial state $\psi$. Then, for each $n$, there exists a QSE protocol 
\begin{equation}
\mathcal{E}_n\colon
\mathcal{L}\left( A^{\otimes n}B^{\otimes n} \otimes D_nE_n\right)
\longrightarrow
\mathcal{L}\left( A^{\otimes n}B^{\otimes n} \otimes F_nG_n\right)
\end{equation}
such that 
\begin{eqnarray}
\left\|
\left( \mathcal{E}_n \otimes \mathrm{id}_{R^{\otimes n}} \right)
\left[ \psi^{\otimes n} \otimes \Psi_n \right]
- 
\psi_\mathrm{f}^{\otimes n} \otimes \Phi_n
\right\|_{1}
&\le&
\epsilon_n^{(2)}, \label{eq:Error5} \\
\lim_{n\to\infty} \frac{1}{n}\left( \log \mathrm{Sr}[\Psi_n] - \log \mathrm{Sr}[\Phi_n] \right) &=& r', \\
\lim_{n\to\infty}\epsilon_n^{(2)}&=&0.
\end{eqnarray}

We need five steps to convert the QSE protocol $\mathcal{E}_n$ to a third-stage protocol $\mathcal{\bar{E}}_n$ of the RAE strategy (see Eq.~(\ref{eq:ErrorRn})):
\begin{equation}
\psi^\mathrm{dec}_n
\xrightarrow[\quad\text{Referee comes back in }\quad]{\quad\mathrm{(i)}\quad}
\psi^\mathrm{dec}_n
\xrightarrow[\quad\text{3-party }\Lambda_n^{- 1}\quad]{\quad\mathrm{(ii)}\quad}
\psi^{\otimes n}
\xrightarrow[\quad\text{2-party }\mathcal{E}_n\quad]{\quad\mathrm{(iii)}\quad}
\psi_\mathrm{f}^{\otimes n}
\xrightarrow[\quad\text{3-party }\Lambda_{n,\mathrm{f}}\quad]{\quad\mathrm{(iv)}\quad}
\psi^\mathrm{dec}_{n,\mathrm{f}}
\xrightarrow[\quad\text{Referee leaves}\quad]{\quad\mathrm{(v)}\quad}
\psi^\mathrm{dec}_{n,\mathrm{f}}.
\end{equation}

(i) For each $n$, Alice and Bob transmit back the parts $R_1^{\otimes n}R_3^{\otimes n}$ and $R_2^{\otimes n}R_4^{\otimes n}$ of the decomposed state $\psi_{n,\mathrm{f}}^\mathrm{dec}$ to the referee.

(ii) Since the transformation $\Lambda$ is reversible, for each $n$, there exists a three-party LOCC
\begin{equation}
\Lambda_n^{- 1}\colon
\mathcal{L}\left(
A_2^{ \otimes \lfloor r_2 n \rfloor}A_1^{ \otimes \lfloor r_1 n \rfloor}A_3^{ \otimes \lfloor r_4 n \rfloor}
 \otimes
B_1^{ \otimes \lfloor r_3 n \rfloor}B_2^{ \otimes \lfloor r_2 n \rfloor}B_3^{ \otimes \lfloor r_4 n \rfloor}
 \otimes
R_1^{ \otimes \lfloor r_1 n \rfloor}R_2^{ \otimes \lfloor r_3 n \rfloor}R_3^{ \otimes \lfloor r_4 n \rfloor}R_4^{ \otimes \lfloor r_4 n \rfloor}
\right)
\longrightarrow
\mathcal{L}\left(
A^{\otimes n} \otimes B^{\otimes n} \otimes R^{\otimes n}
\right)
\end{equation}
such that
\begin{eqnarray}
\left\|
\Lambda_n^{- 1} [ \psi_n^\mathrm{dec} ]
- 
\psi^{\otimes n}
\right\|_{1}
&\le&
\epsilon_n^{(1)}, \label{eq:Error4} \\
\lim_{n\to\infty}\epsilon_n^{(1)}&=&0.
\end{eqnarray}
Alice, Bob, and the referee apply LOCC $\Lambda_n^{- 1}$ to recover the decomposed state to the initial state. 

(iii) For each $n$, Alice and Bob apply the QSE protocol $\mathcal{E}_n$ to their state $\psi^{\otimes n}$.

(iv) The three parties apply the transformation
\begin{equation}
\Lambda_{n,\mathrm{f}}
\colon
\mathcal{L}\left(
A^{\otimes n} \otimes B^{\otimes n} \otimes R^{\otimes n}
\right)
\longrightarrow
\mathcal{L}\left(
A_2^{ \otimes \lfloor r_2 n \rfloor}A_1^{ \otimes \lfloor r_3 n \rfloor}A_3^{ \otimes \lfloor r_4 n \rfloor}
 \otimes
B_1^{ \otimes \lfloor r_1 n \rfloor}B_2^{ \otimes \lfloor r_2 n \rfloor}B_3^{ \otimes \lfloor r_4 n \rfloor}
 \otimes
R_1^{ \otimes \lfloor r_1 n \rfloor}R_2^{ \otimes \lfloor r_3 n \rfloor}R_3^{ \otimes \lfloor r_4 n \rfloor}R_4^{ \otimes \lfloor r_4 n \rfloor}
\right)
\end{equation}
such that
\begin{eqnarray}
\left\|
\Lambda_{n,\mathrm{f}} [ \psi_\mathrm{f}^{\otimes n} ]
- 
\psi_{n,\mathrm{f}}^\mathrm{dec}
\right\|_{1}
&\le&
\epsilon_n^{(3)}, \label{eq:Error6} \\
\lim_{n\to\infty}\epsilon_n^{(3)}&=&0,
\end{eqnarray}
where the role of Alice (Bob) in $\Lambda_{n,\mathrm{f}}$ is the role of Bob (Alice) in $\Lambda_n$, and the role of the referee in $\Lambda_{n,\mathrm{f}}$ is the same as that of the referee in $\Lambda_n$.

(v) Lastly, the referee transmits the parts $R_1^{\otimes n}R_3^{\otimes n}$ and $R_2^{\otimes n}R_4^{\otimes n}$ of $\psi_{n,\mathrm{f}}^\mathrm{dec}$ to Alice and Bob, respectively.

The first and last steps are just a redistribution of quantum states.
Due to the assumptions regarding the referee in the RAE task, these steps are permissible, and the entanglement used in these steps does not affect the achievable entanglement rate $r'$.
We do not explicitly indicate the quantum channels for these. As a result, the five steps are represented as a protocol 
\begin{equation}
\mathcal{\bar{E}}_n[\rho]=
\left(
\left( \Lambda_{n,\mathrm{f}} \otimes \mathrm{id}_{F_nG_n} \right)
\circ
\left( \mathcal{E}_n \otimes \mathrm{id}_{R^{\otimes n}} \right)
\circ
\left( \Lambda_n^{- 1} \otimes \mathrm{id}_{D_nE_n} \right)
\right)[\rho].
\end{equation}
Then, the protocols $\mathcal{\bar{E}}_n$ become the third-stage protocols whose achievable entanglement rate is identical to that of the QSE protocols $\mathcal{E}_n$, i.e.,
\begin{eqnarray}
&&\left\|
\mathcal{\bar{E}}_n \left[ \psi_n^\mathrm{dec} \otimes \Psi_n \right]
- 
\psi_{n,\mathrm{f}}^\mathrm{dec} \otimes \Phi_n
\right\|_{1} \\
&&=
\left\|
\left( \Lambda_{n,\mathrm{f}} \otimes \mathrm{id}_{F_nG_n} \right)
\left[\left( \mathcal{E}_n \otimes \mathrm{id}_{R^{\otimes n}} \right) \left[ \Lambda_n^{- 1}[\psi_n^\mathrm{dec}] \otimes \Psi_n \right] \right]
- 
\psi_{n,\mathrm{f}}^\mathrm{dec} \otimes \Phi_n
\right\|_{1} \\
&&\le
\left\|
\left( \Lambda_{n,\mathrm{f}} \otimes \mathrm{id}_{F_nG_n} \right)
\left[\left( \mathcal{E}_n \otimes \mathrm{id}_{R^{\otimes n}} \right) \left[ \Lambda_n^{- 1}[\psi_n^\mathrm{dec}] \otimes \Psi_n \right] \right]
- 
\Lambda_{n,\mathrm{f}} \left[\psi_\mathrm{f}^{\otimes n} \right] \otimes \Phi_n
\right\|_{1}
+ 
\left\|
\Lambda_{n,\mathrm{f}} \left[\psi_\mathrm{f}^{\otimes n} \right] \otimes \Phi_n
- 
\psi_{n,\mathrm{f}}^\mathrm{dec} \otimes \Phi_n
\right\|_{1} \\
&&\le
\left\|
\left( \Lambda_{n,\mathrm{f}} \otimes \mathrm{id}_{F_nG_n} \right)
\left[\left( \mathcal{E}_n \otimes \mathrm{id}_{R^{\otimes n}} \right) \left[ \Lambda_n^{- 1}[\psi_n^\mathrm{dec}] \otimes \Psi_n \right] \right]
- 
\Lambda_{n,\mathrm{f}} \left[\psi_\mathrm{f}^{\otimes n} \right] \otimes \Phi_n
\right\|_{1}
+ \epsilon_n^{(3)} \\
&&\le
\left\|
\left( \mathcal{E}_n \otimes \mathrm{id}_{R^{\otimes n}} \right) \left[ \Lambda_n^{- 1}[\psi_n^\mathrm{dec}] \otimes \Psi_n \right]
- 
\psi_\mathrm{f}^{\otimes n} \otimes \Phi_n
\right\|_{1}
+ \epsilon_n^{(3)} \\
&&\le
\left\|
\left( \mathcal{E}_n \otimes \mathrm{id}_{R^{\otimes n}} \right) \left[ \Lambda_n^{- 1}[\psi_n^\mathrm{dec}] \otimes \Psi_n \right]
- \left( \mathcal{E}_n \otimes \mathrm{id}_{R^{\otimes n}} \right) \left[ \psi^{\otimes n} \otimes \Psi_n \right]
\right\|_{1}
+ 
\left\|
\left( \mathcal{E}_n \otimes \mathrm{id}_{R^{\otimes n}} \right) \left[ \psi^{\otimes n} \otimes \Psi_n \right]
- 
\psi_\mathrm{f}^{\otimes n} \otimes \Phi_n
\right\|_{1}
+ \epsilon_n^{(3)} \\
&&\le
\left\|
\left( \mathcal{E}_n \otimes \mathrm{id}_{R^{\otimes n}} \right) \left[ \Lambda_n^{- 1}[\psi_n^\mathrm{dec}] \otimes \Psi_n \right]
- \left( \mathcal{E}_n \otimes \mathrm{id}_{R^{\otimes n}} \right) \left[ \psi^{\otimes n} \otimes \Psi_n \right]
\right\|_{1}
+ \epsilon_n^{(2)}+ \epsilon_n^{(3)} \\
&&\le
\left\|
\Lambda_n^{- 1} [\psi_n^\mathrm{dec}] - \psi^{\otimes n}
\right\|_{1}
+ \epsilon_n^{(2)}+ \epsilon_n^{(3)} \\
&&\le
\epsilon_n^{(1)}+ \epsilon_n^{(2)}+ \epsilon_n^{(3)},
\end{eqnarray}
where the inequalities come from the triangle property and the monotonicity of the trace distance~\cite{Wilde2013}. So, its achievable entanglement rate is given by
\begin{equation}
r[\Lambda,\mathcal{\bar{E}}_n] = r',
\end{equation}
which means that any achievable entanglement rate of the QSE task is also that of the RAE task.
Thus, Eq.~(\ref{eq:app:Need}) is proven.

From Eqs.~(\ref{eq:app:RAENeed}) and~(\ref{eq:app:Need}), Theorem~\ref{thm:Converse} is proven, i.e., $l_\mathrm{new}[\Lambda]$ is a converse bound on the QUI $\Upsilon$.

\section{Details for three-party scenario} \label{app:Three}

\begin{figure}[t]
\includegraphics[clip,width=.85\columnwidth]{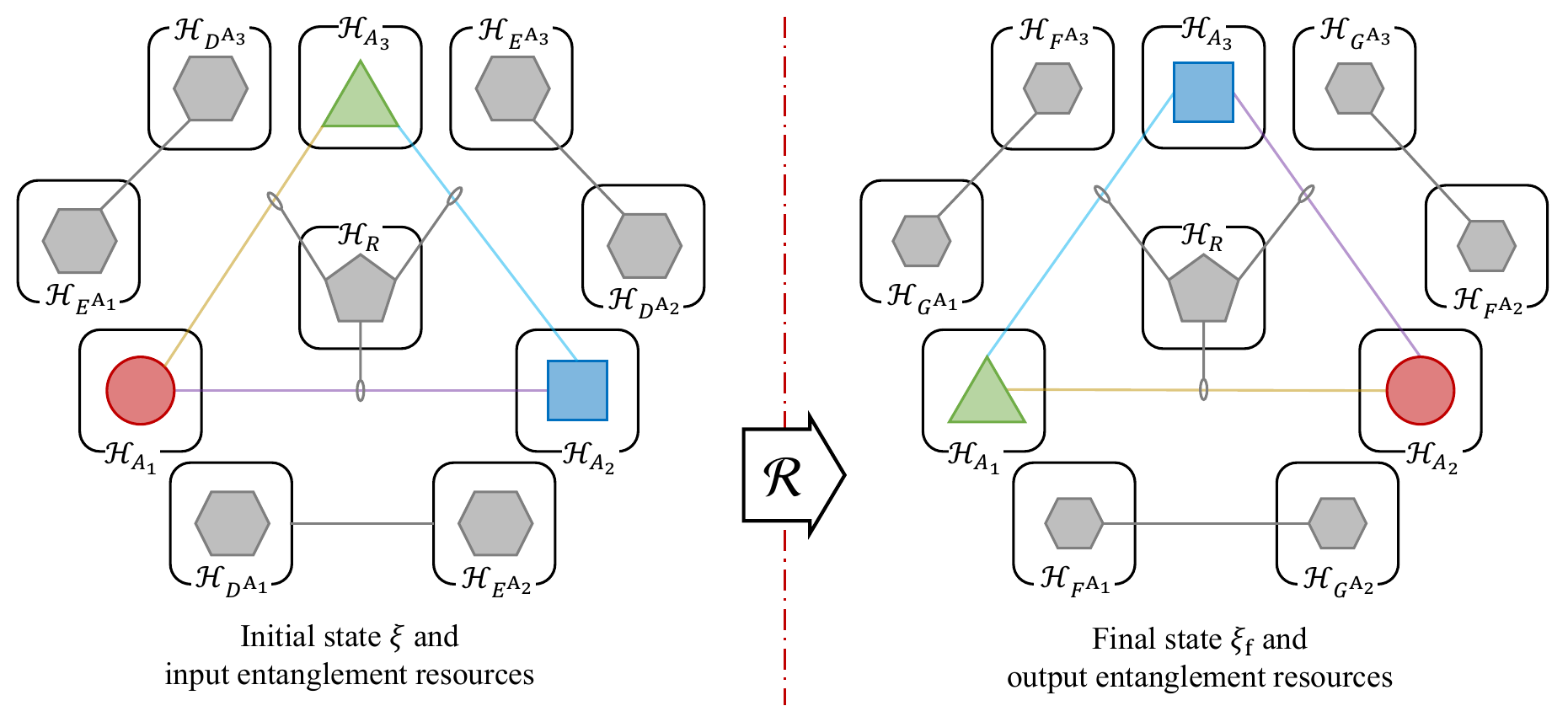}
\caption{
Illustration of QSR protocols:
A QSR protocol $\mathcal{R}$ in Eq.~(\ref{eq:trQSR}) transforms an initial state $\xi$ into a final state $\xi_\mathrm{f}$ in Eq.~(\ref{eq:XiFinal}) by transmitting the quantum state of the $i$th party to the $(i+ 1)$th party for each $i$, while preserving correlations between the reference system $R$ and the others. The three parties utilize bipartite entanglement as non-local resources during the protocol.
}
\label{fig:QSR}
\end{figure}

To describe a QSR protocol, we consider a three-party scenario~\cite{Lee2021} in which any pair of three parties may share bipartite entanglement and apply LOCC.
Let $\xi$ denote an initial state of the QSR task representing quantum systems $A_1A_2A_3R$, where $A_i$ is the quantum system of the $i$th party.
The final state of the QSR task is presented in Eq.~(\ref{eq:XiFinal}).
We assume that the $i$th party holds additional systems $D^{\mathrm{A}_i}E^{\mathrm{A}_i}F^{\mathrm{A}_i}G^{\mathrm{A}_i}$. A quantum channel
\begin{eqnarray}
\mathcal{R}
\colon
\mathcal{L}\left( A_1A_2A_3 \otimes D^{\mathrm{A}_1}E^{\mathrm{A}_2} \otimes D^{\mathrm{A}_2}E^{\mathrm{A}_3} \otimes D^{\mathrm{A}_3}E^{\mathrm{A}_1} \right)
\longrightarrow
\mathcal{L}\left( A_1A_2A_3 \otimes F^{\mathrm{A}_1}G^{\mathrm{A}_2} \otimes F^{\mathrm{A}_2}G^{\mathrm{A}_3} \otimes F^{\mathrm{A}_3}G^{\mathrm{A}_1} \right)
\end{eqnarray}
is a \emph{QSR protocol} with error $\epsilon$ when it can realize the transformation, see Fig.~\ref{fig:QSR},
\begin{eqnarray} \label{eq:trQSR}
\left\|
\left( \mathcal{R} \otimes \mathrm{id}_{R} \right)
\left[ \xi \otimes \Psi^{\mathrm{A}_1\mathrm{A}_2} \otimes \Psi^{\mathrm{A}_2\mathrm{A}_3} \otimes \Psi^{\mathrm{A}_3\mathrm{A}_1} \right]
- 
\xi_\mathrm{f} \otimes \Phi^{\mathrm{A}_1\mathrm{A}_2} \otimes \Phi^{\mathrm{A}_2\mathrm{A}_3} \otimes \Phi^{\mathrm{A}_3\mathrm{A}_1}
\right\|_{1}
\le
\epsilon,
\end{eqnarray}
where $\Psi^{\mathrm{XY}}$ and $\Phi^{\mathrm{XY}}$ denote pure maximally entangled states of the additional systems $D^\mathrm{X}E^\mathrm{Y}$ and $F^\mathrm{X}G^\mathrm{Y}$, respectively. The index XY can be replaced by $\mathrm{A}_1\mathrm{A}_2$, $\mathrm{A}_2\mathrm{A}_3$, or $\mathrm{A}_3\mathrm{A}_1$.

We consider the QSR protocol in the asymptotic regime, i.e., the three parties have $n$ copies of the initial state $\xi$ and aim to achieve $n$ copies of the final state $\xi_\mathrm{f}$ when LOCC and shared entanglement are provided.
To be precise, for $n$ copies of the initial state, let $\mathcal{R}_n$ denote a QSR protocol 
\begin{eqnarray}
\mathcal{R}_n\colon
\mathcal{L}\left( A_1^{\otimes n}A_2^{\otimes n}A_3^{\otimes n} \otimes D^{\mathrm{A}_1}_nE^{\mathrm{A}_2}_n \otimes D^{\mathrm{A}_2}_nE^{\mathrm{A}_3}_n \otimes D^{\mathrm{A}_3}_nE^{\mathrm{A}_1}_n \right)
\longrightarrow
\mathcal{L}\left( A_1^{\otimes n}A_2^{\otimes n}A_3^{\otimes n} \otimes F^{\mathrm{A}_1}_nG^{\mathrm{A}_2}_n \otimes F^{\mathrm{A}_2}_nG^{\mathrm{A}_3}_n \otimes F^{\mathrm{A}_3}_nG^{\mathrm{A}_1}_n\right)
\end{eqnarray}
such that a transformation in the following is realized 
\begin{eqnarray} 
\left\|
\left( \mathcal{E}_n \otimes \mathrm{id}_{R^{\otimes n}} \right)
[ \xi^{\otimes n} \otimes \Psi^{\mathrm{A}_1\mathrm{A}_2}_n \otimes \Psi^{\mathrm{A}_2\mathrm{A}_3}_n \otimes \Psi^{\mathrm{A}_3\mathrm{A}_1}_n ]
- 
\xi_\mathrm{f}^{\otimes n} \otimes \Phi^{\mathrm{A}_1\mathrm{A}_2}_n \otimes \Phi^{\mathrm{A}_2\mathrm{A}_3}_n \otimes \Phi^{\mathrm{A}_3\mathrm{A}_1}_n
\right\|_{1}
\le
\epsilon_n,
\end{eqnarray}
where $\Psi^{\mathrm{XY}}_n$ and $\Phi^{\mathrm{XY}}_n$ denote pure maximally entangled states of the additional systems $D^\mathrm{X}_nE^\mathrm{Y}_n$ and $F^\mathrm{X}_nG^\mathrm{Y}_n$, respectively.
The error $\epsilon_n$ converges to $0$ as $n$ tends to infinity.

A real number $r$ is said to be an \emph{achievable} entanglement rate of the QSR task if the following three limits converge, and the sum of the limits becomes $r$, i.e.,
\begin{eqnarray}
\lim_{n\to\infty} \frac{\log \mathrm{Sr}\left[\Psi^{\mathrm{A}_1\mathrm{A}_2}_n \right] - \log \mathrm{Sr}\left[\Phi^{\mathrm{A}_1\mathrm{A}_2}_n \right]}{n}
&=& r^{\mathrm{A}_1\mathrm{A}_2}, \\
\lim_{n\to\infty} \frac{\log \mathrm{Sr}\left[\Psi^{\mathrm{A}_2\mathrm{A}_3}_n \right] - \log \mathrm{Sr}\left[\Phi^{\mathrm{A}_2\mathrm{A}_3}_n \right]}{n}
&=& r^{\mathrm{A}_2\mathrm{A}_3}, \\
\lim_{n\to\infty} \frac{\log \mathrm{Sr}\left[\Psi^{\mathrm{A}_3\mathrm{A}_1}_n \right] - \log \mathrm{Sr}\left[\Phi^{\mathrm{A}_3\mathrm{A}_1}_n \right]}{n}
&=& r^{\mathrm{A}_3\mathrm{A}_1}, \\
r^{\mathrm{A}_1\mathrm{A}_2} + r^{\mathrm{A}_2\mathrm{A}_3} + r^{\mathrm{A}_3\mathrm{A}_1}
&=& r.
\end{eqnarray}
The minimal achievable entanglement rate over all QSR protocols is defined as
\begin{eqnarray}
\Upsilon^{\mathrm{QSR}}(A_1;A_2;A_3)_\xi=
\inf\left\{r : r \text{ is the achievable entanglement rate of the QSR task for } \xi \right\}.
\end{eqnarray}

\begin{Def}[Three-party common subspace for QSR] \label{def:ThreePartyCommon}
Let $\xi$ be an initial state of the QSR task representing quantum systems $A_1A_2A_3R$ with $\mathcal{H}_{A_1}=\mathcal{H}_{A_2}=\mathcal{H}_{A_3}$.
Let $C^\text{3-com}$ denote a non-empty subspace of the Hilbert spaces.
The subspace $C^\text{3-com}$ is said to be \emph{three-party common} with respect to $\xi$, if there exist \emph{three-party common} unitaries $V_{\mathrm{A}_1}$, $V_{\mathrm{A}_2}$, and $V_{\mathrm{A}_3}$, decomposing $\xi$ into a \emph{three-party common} state $\xi^\text{3-com}$ and a \emph{three-party uncommon} $\xi^\text{3-unc}$, such that
\begin{eqnarray}
\left( V_{\mathrm{A}_1} \otimes V_{\mathrm{A}_2} \otimes V_{\mathrm{A}_3} \otimes \mathds{1}_{R} \right)
\ket{\xi}_{A_1A_2A_3R}
&=& \ket{\xi^\text{3-com}}_{A_1A_2A_3R} + \ket{\xi^\text{3-unc}}_{A_1A_2A_3R}, \\
\ket{\xi^\text{3-com}}_{A_1A_2A_3R}
&=& \ket{\xi^\text{3-com}_\mathrm{f}}_{A_1A_2A_3R}.
\end{eqnarray}
These states are defined as
\begin{eqnarray}
\ket{\xi^\text{3-com}}_{A_1A_2A_3R}
&\coloneqq&
\left( \Pi_{C_{\mathrm{A}_1}^\text{3-com}} V_{\mathrm{A}_1} \otimes \Pi_{C_{\mathrm{A}_2}^\text{3-com}} V_{\mathrm{A}_2} \otimes \Pi_{C_{\mathrm{A}_3}^\text{3-com}} V_{\mathrm{A}_3} \otimes\mathds{1}_{R} \right) \ket{\xi}_{A_1A_2A_3R}, \\
\ket{\xi^\text{3-unc}}_{A_1A_2A_3R}
&\coloneqq&
\left( \Pi_{C_{\mathrm{A}_1}^\perp} V_{\mathrm{A}_1} \otimes \Pi_{C_{\mathrm{A}_2}^\perp} V_{\mathrm{A}_2} \otimes \Pi_{C_{\mathrm{A}_3}^\perp} V_{\mathrm{A}_3} \otimes\mathds{1}_{R} \right) \ket{\xi}_{A_1A_2A_3R},
\end{eqnarray}
where $\xi^\text{3-com}_\mathrm{f}$ denotes the state-rotated state of the three-party common state, the subspace $C^\text{3-com}$ of the Hilbert space $\mathcal{H}_{A_i}$ is denoted by $C_{\mathrm{A}_i}^\text{3-com}$, and $C_{\mathrm{A}_i}^\perp$ is the orthogonal complement of the subspace $C_{\mathrm{A}_i}^\text{3-com}$.
\end{Def}

We present a \emph{subspace rotation} strategy for the QSR task based on a three-party common subspace of an initial state.
In the strategy, three parties state-rotate a three-party uncommon state while leaving a three-party common one.
A subspace rotation protocol consists of five steps as follows:
\begin{eqnarray}
\xi
\xrightarrow[V_{\mathrm{A}_1} \otimes V_{\mathrm{A}_2} \otimes V_{\mathrm{A}_3}]{\quad\mathrm{(i)}\quad}
\left( \xi^{\text{3-com}} + \xi^{\text{3-unc}}\right)
\xrightarrow[U \otimes U \otimes U]{\quad\mathrm{(ii)}\quad}
\xi^{\text{3-str}}
\xrightarrow[\quad\text{Rotating subspaces}\quad]{\quad\mathrm{(iii)}\quad}
\xi^{\text{3-str}}_\mathrm{f}
\xrightarrow[U^\dagger \otimes U^\dagger \otimes U^\dagger]{\quad\mathrm{(iv)}\quad}
\left( \xi^{\text{3-com}} + \xi^{\text{3-unc}}_\mathrm{f}\right)
\xrightarrow[V_{\mathrm{A}_3}^\dagger \otimes V_{\mathrm{A}_1}^\dagger \otimes V_{\mathrm{A}_2}^\dagger]{\quad\mathrm{(v)}\quad}
\xi_\mathrm{f}.
\end{eqnarray}

(i) In the first step, the three parties apply the three-party common unitaries of Definition~\ref{def:ThreePartyCommon}.
They obtain a three-party common state $\xi^\text{3-com}$ and a three-party uncommon state $\xi^{\text{3-unc}}$ in the sense that
\begin{eqnarray}
\ket{\xi}_{A_1A_2A_3R}
\xrightarrow{V_{\mathrm{A}_1} \otimes V_{\mathrm{A}_2} \otimes V_{\mathrm{A}_3}}
\ket{\xi^\text{3-com}}_{A_1A_2A_3R} + \ket{\xi^\text{3-unc}}_{A_1A_2A_3R}.
\end{eqnarray}
Since the parts $A_i$ of the three-party common state $\xi^\text{3-com}$ have already been rotated, it suffices to rotate the parts of the three-party uncommon state $\psi^\text{3-unc}$.

(ii) To separate them, the three parties consider additional systems $A'_i$ and apply the local unitaries $U_{XX'}$ of Eq.~(\ref{eq:UnitaryU}) that pull their uncommon parts towards the systems $A'_1A'_2A'_3$. Then, the uncommon state $\xi^\text{3-unc}$ is stretched in the sense that
\begin{eqnarray}
&&
\left( \ket{\xi^\text{3-com}}_{A_1A_2A_3R} + \ket{\xi^\text{3-unc}}_{A_1A_2A_3R} \right) \otimes\ket{\zeta}_{A'_1} \otimes\ket{\zeta}_{A'_2} \otimes\ket{\zeta}_{A'_3} \\
&&
\xrightarrow{U_{A_1A'_1} \otimes U_{A_2A'_2} \otimes U_{A_3A'_3}}
\ket{\xi^\text{3-str}}
\coloneqq
\ket{\xi^\text{3-com}}_{A_1A_2A_3R} \otimes\ket{\zeta}_{A'_1} \otimes\ket{\zeta}_{A'_2} \otimes\ket{\zeta}_{A'_3} + \ket{\eta}_{A_1} \otimes\ket{\eta}_{A_2} \otimes\ket{\eta}_{A_3} \otimes\ket{\xi^\text{3-unc}}_{RA'_1A'_2A'_3},
\end{eqnarray}
where $\zeta$ and $\eta$ denote any pure states contained in the three-party common subspace and its orthogonal complement, respectively.
We call the resulting state a \emph{three-party stretched} state, denoted by $\xi^\text{3-str}$.

(iii) Thirdly, they state-rotate their respective states in the three-party common subspaces, and they may employ the rest as QSI.
For this, they can utilize the merge-and-merge strategy~\cite{Lee2019a} of Eq.~(\ref{eq:SubspaceRotationStrategy}).
As a result, they rotate the parts $A'_i$ of the stretched state $\xi^\text{3-str}$ as follows:
\begin{equation}
\ket{\xi^\text{3-str}}_{A_1A_2A_3RA'_1A'_2A'_3}\xrightarrow{\text{merge-and-merge}}
\ket{\xi^\text{3-str}_\mathrm{f}}
\coloneqq\ket{\xi^\text{3-com}}_{A_1A_2A_3R} \otimes\ket{\zeta}_{A'_1} \otimes\ket{\zeta}_{A'_2} \otimes\ket{\zeta}_{A'_3} + \ket{\eta}_{A_1} \otimes\ket{\eta}_{A_2} \otimes\ket{\eta}_{A_3} \otimes\ket{\xi^\text{3-unc}_\mathrm{f}}_{RA'_1A'_2A'_3}.
\end{equation}

(iv) The three parties apply the inverse of the unitary $U$, respectively, to put the three-party common and uncommon states together in the sense that
\begin{equation}
\ket{\xi^\text{3-str}_\mathrm{f}}_{A_1A_2A_3RA'_1A'_2A'_3}
\xrightarrow{U^\dagger_{A_1A'_1} \otimes U^\dagger_{A_2A'_2} \otimes U^\dagger_{A_3A'_3}}
\left( \ket{\xi^\text{3-com}}_{A_1A_2A_3R} + \ket{\xi^\text{3-unc}_\mathrm{f}}_{A_1A_2A_3R} \right)
\otimes\ket{\zeta}_{A'_1} \otimes\ket{\zeta}_{A'_2} \otimes\ket{\zeta}_{A'_3}.
\end{equation}

(v) As the last step, they apply the inverses of the three-party common unitaries to obtain the final state $\xi_\mathrm{f}$, i.e.,
\begin{equation}
\ket{\xi^\text{3-com}}_{A_1A_2A_3R} + \ket{\xi^\text{3-unc}_\mathrm{f}}_{A_1A_2A_3R}
\xrightarrow{V_{\mathrm{A}_3}^\dagger \otimes V_{\mathrm{A}_1}^\dagger \otimes V_{\mathrm{A}_2}^\dagger}
\ket{\xi_\mathrm{f}}_{A_1A_2A_3R}.
\end{equation}
Consequently, they complete the QSR task through the subspace rotation strategy.
In the strategy, only the merge-and-merge protocol of Eq.~(\ref{eq:SubspaceRotationStrategy}) consume shared entanglement among the three parties.
So, the net entanglement required in the strategy becomes an achievable bound on $\Upsilon^{\mathrm{QSR}}$.
When the $i$th party starts the merge-and-merge protocol by merging the part $A'_i$ to the next party, the corresponding achievable entanglement rate $v_i^\mathrm{QSR}$ is presented in Eq.~(\ref{eq:NewRateOfQSR}).

In Sec.~\ref{sec:QSR}, we demonstrate that the subspace rotation strategy can be more efficient than the previous one because it needs less bipartite entanglement among the three parties, i.e.,
\begin{equation}
v_\mathrm{new}^\mathrm{QSR} \le u_\mathrm{old}^\mathrm{QSR}.
\end{equation}
To facilitate comparison, we consider the initial state $\xi$ given in Eq.~(\ref{eq:Xi}) and parametrize it as shown in Eq.~(\ref{eq:Parameter}).
For the parametrized state, the achievable entanglement rates $u_i^\mathrm{QSR}$ and $v_i^\mathrm{QSR}$ in Eqs.~(\ref{eq:OldRateOfQSR}) and~(\ref{eq:NewRateOfQSR}) are calculated as
\begin{eqnarray}
u_1^\mathrm{QSR}
&=& - c_0^2\log \frac{c_0^2}{2} - c_1^2\log c_1^2 + 2c_1^2 - c_2^2\log \frac{c_2^2}{2} - c_3^2\log c_3^2, \\
u_2^\mathrm{QSR}
&=& - 2c_0^2\log \frac{c_0^2}{2} + (c_0^2+ c_1^2)\log\frac{c_0^2+ c_1^2}{2} - 2c_1^2\log \frac{c_1^2}{2} - c_2^2\log \frac{c_2^2}{2} - c_3^2\log c_3^3, \\
u_3^\mathrm{QSR}
&=& - c_0^2\log c_0^2 + 2c_0^2 - c_1^2\log \frac{c_1^2}{2} - c_2^2\log \frac{c_2^2}{2} - c_3^2\log c_3^2, \\
v_1^\mathrm{QSR}
&=& - c_0^2\log \frac{c_0^2}{2} + (c_0^2+ c_1^2)\log(c_0^2+ c_1^2) - c_1^2\log c_1^2 + 2c_1^2, \\
v_2^\mathrm{QSR}
&=& - 2c_0^2\log \frac{c_0^2}{2} + 2(c_0^2+ c_1^2)\log(c_0^2+ c_1^2)- (c_0^2+ c_1^2) - 2c_1^2\log \frac{c_1^2}{2}, \\
v_3^\mathrm{QSR}
&=& - c_0^2\log c_0^2 + 2c_0^2 + (c_0^2+ c_1^2)\log(c_0^2+ c_1^2) - c_1^2\log \frac{c_1^2}{2}.
\end{eqnarray}

\end{document}